\documentclass[10pt,a4paper]{article}
\usepackage{jheppub_kim}
\usepackage{pdflscape}
\usepackage{amsmath}
\usepackage{amssymb}
\usepackage{dcolumn}
\usepackage{bm}
\usepackage{color}
\usepackage{epsfig}
\usepackage{amsfonts}
\usepackage{graphicx}
\usepackage{subfigure}
\usepackage{dcolumn}
\usepackage{multirow}
\UseRawInputEncoding

\begin{document}
\title{Thermodynamics of the apparent horizon in the generalized energy-momentum-squared cosmology}

\author[a]{Prabir Rudra,}
\author[b,c]{Behnam Pourhassan}

\affiliation[a] {Department of Mathematics, Asutosh College,
Kolkata-700 026, India.}
\affiliation[b]{School of Physics,
Damghan University, Damghan, 3671641167, Iran.}
\affiliation[c]{Canadian Quantum Research Center 204-3002 32 Ave
Vernon, BC V1T 2L7 Canada.}

\emailAdd{prudra.math@gmail.com, rudra@associates.iucaa.in}
\emailAdd{b.pourhassan@du.ac.ir, b.pourhassan@candqrc.ca}

\abstract{In this note, we explore the thermodynamic properties of
the universe in the background of the generalized
energy-momentum-squared gravity. We derive the energy density of
matter from the non-standard continuity equation and use it in our
analysis. We consider two types of models depending on the nature
of coupling between curvature and matter and perform thermodynamic
analysis on them using the cosmic apparent horizon. The models are
kept as generic as possible from the mathematical point of view in
order to gain a wide applicability of the work. In this work we
have considered power law and exponential form of models. All the
thermodynamic parameters are expressed in terms of the cosmic
apparent horizon radius and its time derivatives and their time
evolution are studied. By using temperature, heat capacity
analysis and the evolution trend of Helmholtz free energy the
conditions for thermodynamic stability of the models are derived.
It is seen that our stability analysis considerably constrain the
parameter space of the model.}


\maketitle

\section{Introduction}
During the last two decades, the study of cosmology have been
centered around the fact that the rate of expansion of the
universe is actually accelerating \cite{acc1, acc2}. Since the
accelerated expansion of the universe is not an expected
phenomenon, the community is bent on finding a proper reason for
this. From the research that have been already performed, it is
seen that there can be a dual possibility regarding the
explanation of this acceleration phenomenon. The first possibility
is the concept of dark energy (DE), which aims at modifying the
matter content of the universe by introducing exotic components
with negative pressure. It is expected that such components will
violate the energy conditions of the universe. A review on dark
energy may be found in the Ref. \cite{de1}. The other possibility,
which has been presented from time to time is the concept of
modified gravity. Here we actually introduce suitable
modifications to the Einstein's gravity so that the modified
theory incorporates the accelerated expansion of the universe. The
reader is suggested to refer to \cite{mod1, mod2, mod3} for
detailed information on modified gravity theories. It should be
stated over here that both the theories aim at modifying the
Einstein's equation of general relativity (GR) in their own ways.
Moreover, dark energy and modified gravity can be shown to be
equivalent to each other via proper fine tuning.

The simplest and the most popular way of modifying the Einstein's
gravity is by replacing the gravity Lagrangian
$\mathcal{L}_{EH}=R$ of the Einstein-Hilbert action by an
analytical function of the Ricci scalar,
$\mathcal{L}_{f(R)}=f(R)$, thus giving rise to $f(R)$ gravity.
Using $f(R)$ modifications to Einstein's gravity, we can explore
the non-linear effects of the curvature of spacetime. Extensive
reviews on $f(R)$ gravity can be found in Refs. \cite{fr1, fr2}.
Further modifications can be affected in the gravity Lagrangian by
introducing an analytical function of Ricci scalar $R$ and the
matter Lagrangian $L_{m}$, giving rise to $f(R,L_m)$ gravity
\cite{frlm}. From such modifications, the contributions coming from
the matter part of the universe are also taken into account along
with the higher order corrections of the spacetime curvature. A
speciality of these theories is that the particles experience an
extra force in the gravity well, in the direction orthogonal to
the four-velocity. Moreover, due to this extra force the particles
undergo a non-geodesic motion. Further developments in $f(R,
L_{m})$ theories can be found in Refs. \cite{frlm2, frlm3, frlm4}.

Narrowing down on these classes of theories Harko et al in
\cite{harko1} proposed the $f(R,T)$ theory where the matter
Lagrangian is given by the scalar $T$ which represents the trace
of the energy-momentum tensor $T_{\mu\nu}$. Now replacing ordinary
matter by a scalar field we get $f(R,T^{\phi})$ theories
\cite{harko1} where $T^{\phi}$ is the trace of the energy momentum
tensor of a scalar field $\phi$. Further attempts to generalize
such theories resulted in the development of
$f(R,T,R_{\mu\nu}T^{\mu\nu})$ theories \cite{haghani}, where a
coupling between curvature and matter is called into play. From
the Lagrangian it is seen that contributions from contraction
between the curvature tensor and the energy-momentum tensor also
comes into play. So the basic idea is to create new scalar
invariants via contraction of tensors and explore their effects on
the dynamics of the universe. Following this path Katirci and
Kavuk \cite{emsg} proposed the Energy-momentum-squared gravity
(EMSG) where the gravity Lagrangian is given by an analytic
function $f(R,T_{\mu\nu}T^{\mu\nu})$ containing the Ricci scalar
and the contraction between energy momentum tensors. This is a
covariant generalization to GR where we allow a term proportional
to $T_{\mu\nu}T^{\mu\nu}$ to be present in the gravity Lagrangian.

Since its induction, the theory has received very good response
and a fair amount of research has been performed on EMSG.
Cosmology in EMSG theory was studied in \cite{board1, akarsu2}.
Cosmological bouncing scenario to avoid the singularity was
studied by Roshan and Shojai in \cite{roshan1} using a particular
model $f(R, T^2)=R+\eta T^{2}$, where $\eta$ is a constant
representing the coupling parameter. A dynamical system analysis
in the background of EMSG was studied by Bahamonde et. al in
\cite{rudra1}. A generalization of EMSG has been achieved via the
energy-momentum-powered gravity (EMPG) by the authors of the
Refs. \cite{board1, akarsu3}. In this theory they have introduced
the model $f(R,T^2)=R+\eta (T^{2})^{n}$ where $\eta$ is the
coupling parameter and $n$ is the power parameter. Observational
data from neutron stars have been used to constrain model
parameters of EMSG by Akarsu et al in \cite{akarsu1}. Matter
wormhole solutions have been explored in the background of EMSG in
\cite{moraes1}. Other important studies in EMSG gravity can be
found in Ref. \cite{nari1, keskin1}. Continuing the journey of
modifications Akarsu et al. in \cite{akarsu4} proposed the
energy-momentum-Log gravity (EMLG), where a specific form of
logarithmic function $f(T_{\mu\nu}T^{\mu\nu})=\alpha \ln(\lambda
T_{\mu\nu}T^{\mu\nu})$ is considered, $\alpha$ and $\lambda$ being
constants.

It was as late as the 1970s that scientists began to understand
that there is a deep underlying connection between thermodynamics
and gravitation. The initial breakthrough in this topic was
achieved via black holes (BH). It was found that the area of the
horizon of a BH is connected with the entropy of the system. This
presented a direct link between the geometry of the system with
thermodynamics. Motivated from this the initial form of
thermodynamic studies were limited to black hole thermodynamics
\cite{bht1}. It was shown in Ref. \cite{bht2} that the surface
gravity of a BH is related to the temperature, and it was also
shown that these quantities satisfy the first law of
thermodynamics (FLT), $\delta Q = T ds$. Using this FLT and the
properties of BH entropy the authors of \cite{jacob} derived the
Einstein's equation of GR. Friedmann equations were derived from
the thermodynamic point of view by Cai and Kim in \cite{caikim}.
The first law of thermodynamics at the cosmic apparent horizon
using the modified Friedmann equations have been studied for
general relativity, Gauss-Bonnet, and Lovelock gravity
\cite{caikim, caiakbar}, then extended to the scalar-tensor
gravity \cite{6}, braneworld universe \cite{8,9,10},
Horava-Lifshitz gravity \cite{12,13}, $f(R)$ gravity \cite{7},
$f(R,\mathcal{L})$ theory \cite{frlm4} generic $f(R, \phi,
\partial\phi)$ gravity \cite{11}. From the physical point of view,
apparent horizons are related to the observable boundary of the
universe. On the other hand, from mathematical point of view,
there are two kind of apparent horizons. One of them is related to
the black holes, while the other is situated near the expanding
boundary of universe. The latter is called the cosmic apparent
horizon which is not a null surface. Using the Palatini formalism
the laws of thermodynamics were studied in the background of
$f(R)$ gravity in \cite{bamboo}. Thermodynamic prescription of
cosmological horizons in the background of $f(T)$ gravity was
studied in Ref. \cite{bambooft}. Thermodynamics in the background
of $f(R,T)$ gravity theory using the FLRW spacetime was
investigated in \cite{therm1}. The authors of Ref. \cite{frlm4}
studied thermodynamics in the background of $f(R,\mathcal{L})$
theories, where $\mathcal{L}$ represents the matter Lagrangian
density. Charged black hole perturbations in this theory was
studied by the authors in \cite{new1}. The effects of coupling
between matter and geometry components of the universe is a very
important aspect of modern cosmology and quite expectedly
thermodynamics is reasonably sensitive to such couplings. There is
a clear indication of this in the Refs. \cite{therm1, frlm4}.

In this work we would like to further develop the EMSG theory by
performing a thermodynamic study of universe in its background. A
thermodynamic analysis of any cosmological model is very important
as far as its viability as a successful model is concerned.
Precisely any model which has an aim to become a successful model
of universe must satisfy the thermodynamic conditions of the
universe. We would also like to investigate various forms of
coupling effects on the thermodynamic properties of the model. So
the motivation of the work is very straightforward and moreover
this is probably the first attempt towards studying the
thermodynamics of the universe in EMSG theory. The paper is
organized as follows: In section II we have listed the basic
equations of EMSG theory. Section III has been dedicated to
selection of specific models. In section IV a detailed
thermodynamic study is performed. Finally the paper ends with a
discussion and conclusion in section V.

\section{Basic equations of Energy-momentum squared cosmology}

The action of the energy-momentum-squared gravity model is written
as~\cite{emsg, board1, akarsu3}
\begin{equation}\label{action}
S=\frac{1}{2\kappa^2}\int d^4x \sqrt{-g} f(R,\mathbf{T^2})  +
S_{\rm m},
\end{equation}
where $f$ is a function depending on the square of the
energy-momentum tensor $\mathbf{T^2}=T^{\mu\nu}T_{\mu\nu}$ and the
scalar curvature $R$. Here, $\kappa^2=8\pi G$ and $S_{\rm m}$
represents the action corresponding to the matter component.

On varying the action (\ref{action}) with respect to the metric
$g_{\mu\nu}$ we arrive at the following field equations
\begin{equation}
R_{\mu \nu}f_R  +g_{\mu\nu} \Box f_R-\nabla_{\mu}\nabla_{\nu}
f_R-\frac{1}{2} g_{\mu\nu}f=\kappa^2 T_{\mu \nu}-f_{\mathbf{T^2}}
\Theta_{\mu \nu}\,, \label{FieldEq}
\end{equation}
where $\Box=\nabla_\mu \nabla^\mu$, $f_{R}=\partial f/\partial R$,
$f_{\mathbf{T^2}}=\partial f/\partial \mathbf{T^2}$ and
\begin{equation}
    \Theta_{\mu\nu}=\frac{\delta (\mathbf{T^2})}{\delta g^{\mu\nu}}= \frac{\delta (T^{\alpha\beta}T_{\alpha\beta})}{\delta g^{\mu\nu}}=-2L_{\rm m}\Big(T_{\mu\nu}-\frac{1}{2}g_{\mu\nu}T\Big)-T\, T_{\mu\nu}+2T^{\alpha}_{\mu}T_{\nu\alpha}-4T^{\alpha\beta}\frac{\partial^2 L_{\rm m}}{\partial g^{\mu\nu}\partial g^{\alpha\beta}}\,,\label{Theta}
\end{equation}
where $T$ is the trace of the energy-momentum tensor. By taking
covariant derivatives in the field equation~\eqref{FieldEq}, one
finds the following conservation equation
\begin{eqnarray}
\kappa^2\nabla^\mu T_{\mu\nu}=-\frac{1}{2}g_{\mu\nu}\nabla^\mu
f+\nabla^\mu(f_{\mathbf{T}^2}\Theta_{\mu\nu})\,.\label{conservation1}
\end{eqnarray}
As one can see from the above equation that the standard
conservation equation $\dot{\rho}+3H\left(\rho+p\right)=0$ does
not hold for this theory. Instead we will get a non-standard
continuity equation from the above equation which will govern the
properties of matter and the matter energy density of this system.
It should be noted here that this is a unique feature of this
theory and occurs due to the curvature-matter (squared form)
coupling in the gravitational Lagrangian.

In the following, we will concentrate on the flat FLRW cosmology
for this model whose metric is described by
\begin{equation}
ds^2=-dt^2+a^2(t)\delta_{ik}dx^idx^k,
\end{equation}
where $\delta_{ik}$ is the Kronecker delta and $a(t)$ the scale
factor representing the expansion of the universe. Here we will
consider that the matter content is described by a standard
perfect fluid with the energy momentum tensor given by,
$T_{\mu\nu}=(\rho+p)u_\mu u_\nu + p g_{\mu\nu}$ with $u_{\mu}$
being the 4-velocity and $\rho$ and $p$ are the energy density and
the pressure of the fluid respectively. Using this energy-momentum
tensor we get $T^2=T_{\mu\nu}T^{\mu\nu}=\rho^2+3p^2$. Further, let
us assume $L_{\rm m}=p$ which allows us to rewrite
$\Theta_{\mu\nu}$ defined in equation \eqref{Theta} as a quantity which
does not depend on the function $f$, as given below \cite{emsg,
board1}
\begin{equation}\label{bigtheta}
\Theta_{\mu\nu}=-\Big(\rho^2+4 p\rho+3p^2\Big)u_\mu u_\nu\,.
\end{equation}
The modified FLRW equations which corresponds to this particular
action are given by
\begin{eqnarray}
-3f_R\Big(\dot{H}+ H^2\Big)+\frac{f}{2}+3 H \dot{f_R}&=&\kappa^2\Big(\rho+\frac{1}{\kappa^2}f_{\mathbf{T^2}}\mathbf{\Theta^2}\Big)\,,\label{FW1} \\
-f_R(\dot{H}+3 H^2)+\frac{1}{2}f+\ddot{f_R}+2 H
\dot{f_R}&=&-\kappa^2 p\,,\label{FW2}
\end{eqnarray}
where dots denote differentiation with respect to the cosmic time
$t$, $H=\dot{a}/a$ is the Hubble parameter and using
Eq. (\ref{bigtheta}) the expression for $\mathbf{\Theta^2}$ is
calculated as,
\begin{equation}
\mathbf{\Theta^2}\equiv\Theta_{\mu\nu}\Theta^{\mu\nu}=\rho^2+4p
\rho+3p^2\label{Theta2}
\end{equation}
The conservation equation \eqref{conservation1} can be written as
follows
\begin{eqnarray}
\kappa^2\left[\dot{\rho}+3H(\rho+p)\right]&=&-\mathbf{\Theta^2}\dot{f}_{\mathbf{T^2}}-f_{\mathbf{T^2}}
\Big[3 H\mathbf{\Theta^2}+\frac{d}{dt}\Big(2\rho
p+\frac{1}{2}\mathbf{\Theta^2}\Big)\Big]\,.\label{conservation3}
\end{eqnarray}
Clearly as stated above the standard conservation equation does
not hold in $f(R,\mathbf{T^2})$ cosmology for an arbitrary
function. The covariant divergence of the field equations produces
non-zero terms on the right hand side, thus leading to the
modified continuity equation given above. If one chooses
$f(R,\mathbf{T^2})=f(R)$, all the terms on the RHS of the above
equation are zero and the standard conservation equation is
recovered.

We can rewrite the modified FLRW equations in the standard form
as,
\begin{eqnarray}
    3H^2&=&\kappa^{2}\rho_{eff}=\kappa^2(\rho+\rho_{\rm modified})\,,\\
      3H^2+2\dot{H}&=&-\kappa^{2}p_{eff}=-\kappa^2(p+p_{\rm modified})\,,
\end{eqnarray}
where we have defined the energy density and pressure for the EMSG
modifications as
\begin{eqnarray}
\rho_{\rm modified}&=&-\frac{1}{f_R}\left[\rho+\frac{1}{\kappa^{2}}\left\{f_{T^2}\left(\rho^{2}+4p\rho+3p^{2}\right)-\frac{f}{2}-3H\dot{f}_R+3\dot{H}f_{R}\right\}\right]-\rho\label{rho}\\
p_{\rm
modified}&=&-\left[\frac{1}{f_R}\left\{p+\frac{1}{\kappa^{2}}\left(\frac{f}{2}+\ddot{f}_{R}+2H\dot{f}_{R}\right)\right\}+\frac{\dot{H}}{\kappa^{2}}\right]-p\label{p}
\end{eqnarray}
$\rho$ and $p$ are respectively the energy density and pressure of
matter. It should be noted here that the energy density and
pressure contributions from the modified gravity can be considered
equivalent to the contributions from a dark energy component and
so our aim is to consider non-exotic components in the matter
sector. This will facilitate a better understanding of the exotic
nature of the modified gravity. In the following, a standard
barotropic equation of state will be assumed for the matter fluid
as given by,
\begin{equation}\label{pressure1}
p=w \rho
\end{equation}
where $w$ is the equation of state (EoS) parameter. Using this
relation in equation (\ref{Theta2}) one gets,
\begin{equation}\label{theta1}
\mathbf{\Theta^2}=(1+4w+3w^2)\rho^2
\end{equation}
Also the conservation equation \eqref{conservation3} becomes
\begin{eqnarray}
\dot{\rho}+3H(w+1)\rho&=&-f_{\mathbf{T^2}} \left[3 \left(3 w^2+4 w+1\right) H \rho ^2+\left(3 w^2+8 w+1\right) \rho \dot{ \rho}\right]\nonumber\\
&&-\left(3 w^2+4 w+1\right) \rho^2
\dot{f}_{\mathbf{T^2}}\,\label{conservation4}
\end{eqnarray}
Finally we can define the effective equation of state (EoS) as,
\begin{equation}
w_{eff}=\frac{p_{eff}}{\rho_{eff}}= \frac{w\rho+p_{\rm
modified}}{\rho+\rho_{\rm modified}}\,.\label{eos2}
\end{equation}
In order to realize the late cosmic acceleration we should have
$w_{eff}<-1/3$, which corresponds to dark energy. Since we aim to
consider the matter EoS, $w\geq -1/3$ (non-exotic), then the role
of the EoS of modified gravity $w_{modified}$ is so much more
significant for realizing the accelerated expansion of the
universe, where we have considered
$w_{modified}=\frac{p_{modified}}{\rho_{modified}}$.

Since the continuity equation given by equation
(\ref{conservation4}) is non-standard consisting of unorthodox
terms in the RHS, it is not a trivial task to integrate it for
this model. The obvious reason being the non-linear terms of
$\rho$ in the RHS of the equation, which makes it a non-linear
differential equation. The non-zero term on the right hand side of
the modified continuity equation (\ref{conservation4}) poses a
real mathematical challenge for this operation. Here we would like
to solve the continuity equation in a model independent way and
express the energy density parameter $\rho$ in terms of the
redshift parameter $z$. We see that the conservation equation
(\ref{conservation4}) is not integrable for any arbitrary value of
$w$ by the known mathematical methods. This means that we will
have to input numerical values for $w$ in the equation and check
for solutions. From the work of Board et al., in \cite{board1} we
see that the equation is integrable for only $w=-1/3$ and $w=-1$.
We know that $w=-1$ corresponds to the $\Lambda CDM$ cosmology and
$w<-1/3$ indicates the boundary between the exotic and non-exotic
sectors. So from our perspective the solution corresponding to
$w=-1/3$ is crucial. On solving equation \eqref{conservation4} for
$w=-1$, we get two real solutions for the density parameter $\rho$
given by \cite{our1},
\begin{equation}\label{cons1}
\rho=\frac{1}{f_{\mathbf{T^2}}}~~~~~~~~~~~   and
~~~~~~~~~~\rho=C_{0}
\end{equation}
where $C_{0}$ is a constant.\\

For $w=-1/3$ we get only one real value for $\rho$ given by
\cite{our1},
\begin{equation}\label{cons2}
\rho=-\frac{3W\left[\frac{4}{3}\left\{-e^{-C_1}
(f_{\mathbf{T^2}})^3
\left(z+1\right)^6\right\}^{1/3}\right]}{4f_{\mathbf{T^2}}}
\end{equation}
where $W\left[y\right]$ is the Lambert $W$ function (see
appendix), $z$ is the redshift parameter and $C_{1}$ is the
constant of integration. For the reader's convenience we would
like to provide a short mathematical definition of the Lambert $W$
function. Although the expressions of $\rho$ for $w=-1$ are too
trivial yet we will use the values of $\rho$ obtained for both
$w=-1, -1/3$ for our further analysis. But it is to be noted that
the solution generated for $w=-1/3$ will be the one that we will
expect to give us more interesting results. The reason is
straightforward and has already been discussed above. We should
state here that due to the narrow range of solution obtained for
$\rho$ our thermodynamic analysis may be constrained
significantly. But this is a property of the model and needs to be
accepted given the limitations of our mathematical capabilities.
But we should state here that as far as cosmological implication
\cite{our1} is concerned our solution is quite fine and ready to
produce interesting results.

\section{Selection of Model}

In literature some models of EMSG can be found which have shown
promising results till now. In Ref. \cite{board1} Board and Barrow
considered a fairly generic model that gave promising results as
far as cosmology is concerned. The model is given by,
\begin{equation}\label{submodel1}
f(R,T^2)=R+\eta (T^2)^{n}
\end{equation}
where $\eta$ and $n$ are constant parameters. This is a
generalized form of EMSG known as energy-momentum powered gravity
(EMPG) \cite{board1}. Some solutions of this model with $n=1/2$
and $n=1/4$ have been discussed in Ref. \cite{board1}. In the EMPG
model $n>1/2$ corresponds to high energy densities and thus
compatible with early universe. $n<1/2$ correspond to low energy
densities and thus suit the late universe. For $n=1$ this reduces
to the following special case used in Refs. \cite{emsg, roshan1}
\begin{equation}\label{submodel11}
f(R,T^2)=R+\eta T^2
\end{equation}
Using the same functional form a more generic model of EMSG was
considered in Ref. \cite{rudra1} as given below,
\begin{equation}\label{submodel12}
f(R,T^2)=\alpha R^{n}+\beta (T^2)^{m}
\end{equation}
where $\alpha$, $\beta$, $n$ and $m$ are all constants. In
\cite{rudra1} another model of a different form was considered
given by,
\begin{equation}\label{submodel10}
f(R,T^2)=f_{0}R^{n}(T^{2})^{m}
\end{equation}
where $f_{0}$, $n$ and $m$ are constants. In Ref. \cite{akarsu4}
the authors have studied a special class of EMSG models called the
energy-momentum-log gravity (EMLG) which was characterized by the
form $f(T_{\mu\nu}T^{\mu\nu})=\alpha \ln{(\lambda
T_{\mu\nu}T^{\mu\nu})}$. Here $\alpha$ is a constant and $\lambda$
has dimensions inverse energy density squared so that $\lambda
T_{\mu\nu}T^{\mu\nu}$ is dimensionless. This form has some
specific advantageous features as discussed in Ref. \cite{akarsu4}.

Motivated by all the above mentioned models, we proceed to
consider some generic models for the present study. Our idea is to
consider some generic mathematical functions that will help us
explore the effects of the the scalar invariants $R$ and $T^2$ and
their coupling on the thermodynamic properties. We will basically
consider two different types of models and believe that all other
models will be some sub-classes of either of the two forms
considered in the present study. In that sense this work will have
a far wider range compared to the other works.

\subsection{Minimal coupling between $R$ and $T^{2}$}
Here we will consider the models of the form:
$f(R,T^2)=f_{1}(R)+f_{2}(T^2)$, where $f_{1}(R)$ and
$f_{2}(T^{2})$ are analytic functions of $R$ and $T^{2}$
respectively. We can see that here the curvature $R$ and the
matter component $T^{2}$ are coupled minimally in the additive
sense. Now we may generate various toy models by considering
various functional forms for $f_{1}(R)$ and $f_{2}(T^{2})$ along
with coupling constants.

\subsubsection{Power Law models}
We consider, $f_{1}(R)=\alpha_{1} R^{n}$, $f_{2}(T^2)=\alpha_{2}
(T^2)^{m}$. So the model is given by,
\begin{equation}\label{submodel13}
f(R,T^2)=\alpha_{1} R^{n}+\alpha_{2} (T^2)^{m}
\end{equation}
where $\alpha_{1}$, $\alpha_{2}$, $n$ and $m$ are constants. Here
$\alpha_{1}$ and $\alpha_{2}$ act as coupling constants between
the geometric and matter sectors. This model is identical with the
model given in equation (\ref{submodel12}) \cite{rudra1}. This model
comprises of two power law forms, one each on $R$ and $T^2$
combined together additively. We call this model energy-momentum
doubly powered gravity (EMDPG). This model will help us explore
the non-linear effects of the scalar invariants on the
thermodynamic properties. For $\alpha_{1}=n=1$, we get the model
given in equation (\ref{submodel1}). Moreover for $\alpha_{1}=n=m=1$,
we get the model given in equation (\ref{submodel11}). The model
reduces to GR for $\alpha_{1}=n=1$ and $\alpha_{2}=0$.

\subsubsection{Exponential models}
Here we consider $f_{1}(R)=g_{1}\exp{(\beta_{1}R)}$,
$f_{2}(T^2)=g_{2}\exp{(\beta_{2}T^2)}$. So the model becomes,
\begin{equation}\label{submodel14}
f(R,T^2)=g_{1}\exp{(\beta_{1}R)}+g_{2}\exp{(\beta_{2}T^2)}
\end{equation}
where $g_{1}$, $g_{2}$, $\beta_{1}$ and $\beta_{2}$ are constants.
Here $g_{1}$ and $g_{2}$ act as coupling constants. As can be seen
from the model, here we have considered exponential forms for both
$R$ and $T^2$. We name this model energy-momentum doubly
exponential gravity (EMDEG). For $g_{1}=1, g_{2}=0$ and retaining
the linear terms from the Taylor series expansion of the first
exponential we can realize GR from this model.



\subsection{Non-minimal coupling between $R$ and $T^{2}$}
Here the model is characterized by
$f(R,T^2)=f_{1}(R)+f_{2}(R)f_{3}(T^2)$, where $f_{1}(R)$,
$f_{2}(R)$ are analytic functions of $R$ and $f_{3}(T^{2})$ is an
analytic function of $T^{2}$. This is our second form of the
models which will be considered in this study. In this form, two
scalar invariants $R$ and $T^2$ are non-minimally coupled (NMC) to
each other (multiplicative sense). Now we may construct various
toy models by considering specific functional forms for the
analytic functions. Here we will consider one toy model for our
analysis as given below.

\subsubsection{Power law models}
We consider $f_{1}(R)=\alpha_{1} R^{n}$,
$f_{2}(R)=\alpha_{2}R^{m}$, $f_{3}(T^2)=(T^{2})^{l}$. So the model
becomes,
\begin{equation}\label{submodel21}
f(R,T^2)=\alpha_{1} R^{n}+\alpha_{2}R^{m}(T^{2})^{l}
\end{equation}
where $\alpha_{1}$, $n$, $\alpha_{2}$, $m$ and $l$ are constants.
Here $\alpha_{2}$ is the coupling parameter. For $\alpha_{1}=0$ we
get the model discussed in equation (\ref{submodel10}). We name this
model energy-momentum-triply-powered-gravity (EMTPG). The model
reduces to GR for $\alpha_{1}=n=1$ and $\alpha_{2}=0$.
Investigating this model we will try to understand the effects of
non-minimal curvature matter coupling on the thermodynamics of the
universe.


\section{Thermodynamics in EMSG}
In this section we will study the thermodynamics of the universe
in the background of EMSG. Our idea is to investigate various
thermodynamic parameters for the toy models of EMSG as discussed
above. The basic aim of the study will be to check the
thermodynamic stability of the models and thus constrain the
parameter space in such an attempt. In order to study the
thermodynamics of the model we should first express the Ricci
scalar in terms of the Hubble expansion parameter $H$ as given
below \cite{frlm4},
\begin{equation}\label{Ricci2}
R=6(\dot{H}+2H^{2}).
\end{equation}
The Hubble expansion parameter can be written in terms of the
cosmic apparent horizon as,
\begin{equation}\label{r}
H=r_{A}^{-1}
\end{equation}
where $r_{A}$ is the radius of the cosmic apparent horizon. Therefore
equation (\ref{Ricci2}) is reduced to the following relation,
\begin{equation}\label{Ricci3}
R=\frac{6}{r_{A}^{2}}|2-\dot{r}_{A}|.
\end{equation}

As we know, in GR, the horizon entropy is given by the
Bekenstein-Hawking (B-H) formula, which is equal to the apparent
horizon area divided by $4G$, where $G$ is gravitational coupling
\cite{012, 013, 014}. However, in modified gravitational theories
we have the Wald entropy \cite{wald} which is obtained using the
replacement of $G$ by the effective gravitational coupling
($G\rightarrow G_{eff}=G^{\prime}/f_{R}$, where
$G^{\prime}\equiv\mathcal{F}=1+f_{\mathbf{T^2}}/8\pi G$
\cite{Brustein}, so at leading order, one can write
$G^{\prime}\approx G$, which is our case of interest). Hence the
Wald entropy of generalized energy-momentum-squared gravity can be
expressed as following,
\begin{equation}\label{S}
S=\frac{\mathcal{A}f_{R}}{4G},
\end{equation}
where
\begin{equation}\label{A}
\mathcal{A}=4\pi r_{A}^{2}
\end{equation}
is the cosmic apparent horizon area. Similarly the thermodynamic volume of
the system may be given by,
\begin{equation}\label{V}
V=\frac{4}{3}\pi r_{A}^{3}
\end{equation}
Moreover, using the surface gravity ($\kappa_{s}$) at the cosmic apparent
horizon and the equation (\ref{r}) one can obtain the temperature
of the cosmic apparent horizon as \cite{caikim},
\begin{equation}\label{T}
\bar{T}=\frac{\kappa_{s}}{2\pi}=\frac{|1-\frac{\dot{r}_{A}}{2}|}{2\pi r_{A}}.
\end{equation}
Then, one can obtain specific heat at constant volume by using the following general formula,
\begin{equation}\label{C}
C_{V}=\bar{T}\left(\frac{\partial S}{\partial \bar{T}}\right)_{V}.
\end{equation}
$C_{V}$ is a very important quantity to study the model stability
from thermodynamic point of view. We know that if it is negative,
then the system is said to be in an unstable phase, whereas
positivity of $C_{V}$ shows the stability of the system. Obviously
this means that by keeping the volume constant, we are invariably
talking about a closed system. On the other hand, if we think
about the horizon of the expanding universe it is not a closed
system. But it must be stated here that all the thermodynamic
parameters are calculated taking into consideration that the time
is fixed. In such a scenario the changing scale factor or the
expansion does not have any role to play and considering a
constant volume is possible.

Along with this we need to have a positive temperature $T>0$. Our
main goal of this paper is to find time dependent cosmic apparent
horizon by using the thermodynamic rules. Then, we can use it to
determine Hubble expansion parameter and hence scale factor.
Having scale factor we can study all cosmological consequences. An
important cosmological parameter is the deceleration parameter
given by,
\begin{eqnarray}\label{q}
q=-(1+\frac{\dot{H}}{H^{2}}).
\end{eqnarray}
Exploring its evolution for the interesting models described above
will be a straightforward task and by properly fine tuning the
parameters we can realize the accelerated expansion of the
universe. The internal energy could be expressed as \cite{frlm4},
\begin{eqnarray}\label{U}
U=V\rho_{eff}
\end{eqnarray}
and thermodynamical work can be given by,
\begin{eqnarray}\label{W}
W=\frac{\rho_{eff}-p_{eff}}{2}.
\end{eqnarray}
Therefore, one can write the first law of thermodynamics as
\begin{eqnarray}\label{WT}
\bar{T}dS=dU-WdV.
\end{eqnarray}
Finally, the Helmholtz free energy is given by,
\begin{eqnarray}\label{F}
F=U-\bar{T}S.
\end{eqnarray}
This is another parameter which is closely linked to the
thermodynamic stability of a system. It is basically a
thermodynamic potential that measures the useful work obtained
from a closed thermodynamic system at constant temperature. Here
we consider a fixed time to realize this constant temperature and
closed system by eliminating the effects of expanding universe. At
constant temperature $F$ is minimized under equilibrium condition.

Above we have reviewed the basic thermodynamic parameters that
will be used in this study. Now we will proceed to investigate the
evolution of the above discussed parameters in the various EMSG
models separately.

\subsection{Thermodynamics in EMDPG model}

\begin{figure}[h!]
 \begin{center}$
 \begin{array}{cccc}
\includegraphics[width=60 mm]{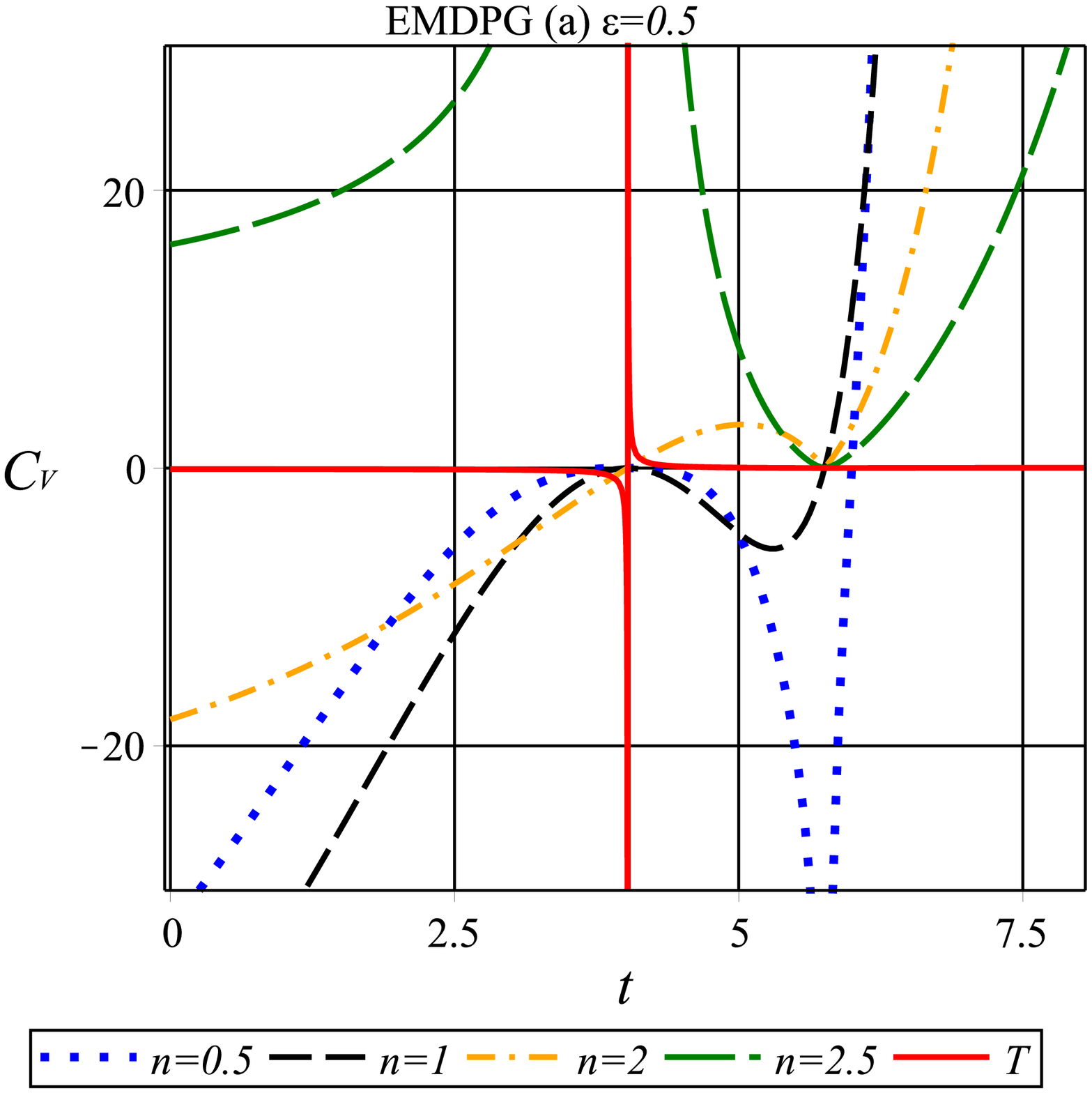}~~~~~~\includegraphics[width=60 mm]{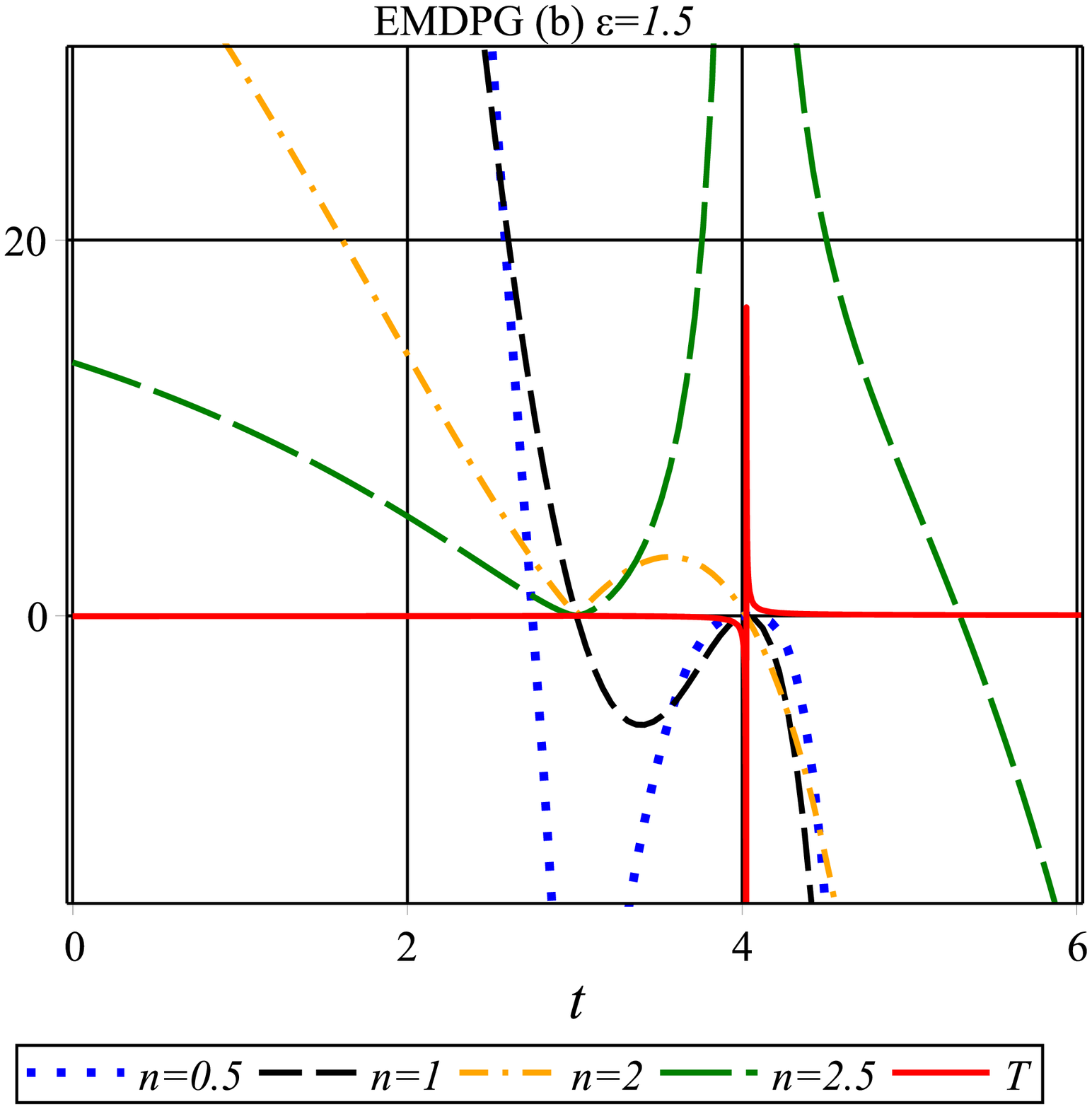}
 \end{array}$
 \end{center}
 \begin{center}$
 \begin{array}{cccc}
\includegraphics[width=60 mm]{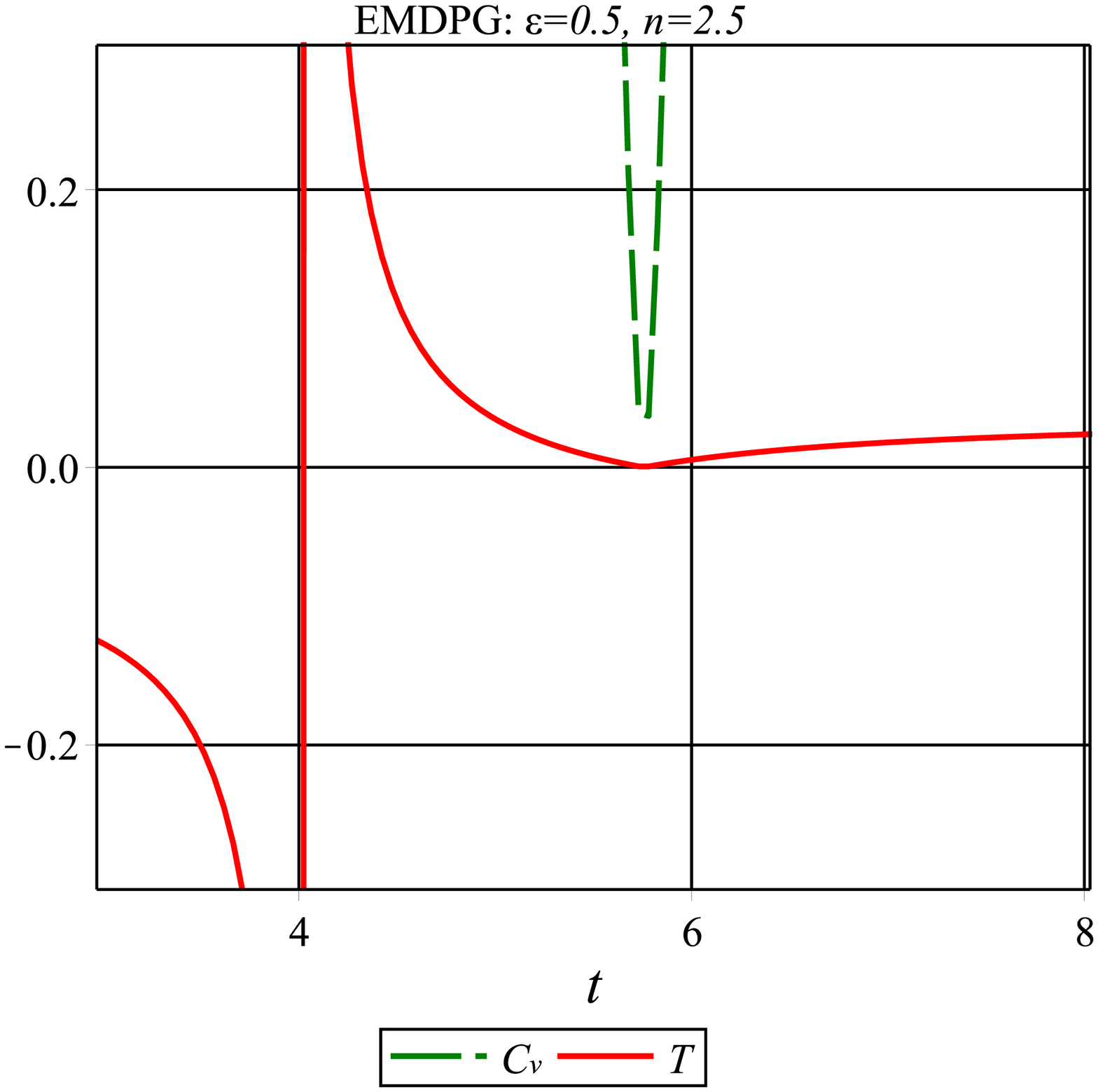}~~~~~~\includegraphics[width=60 mm]{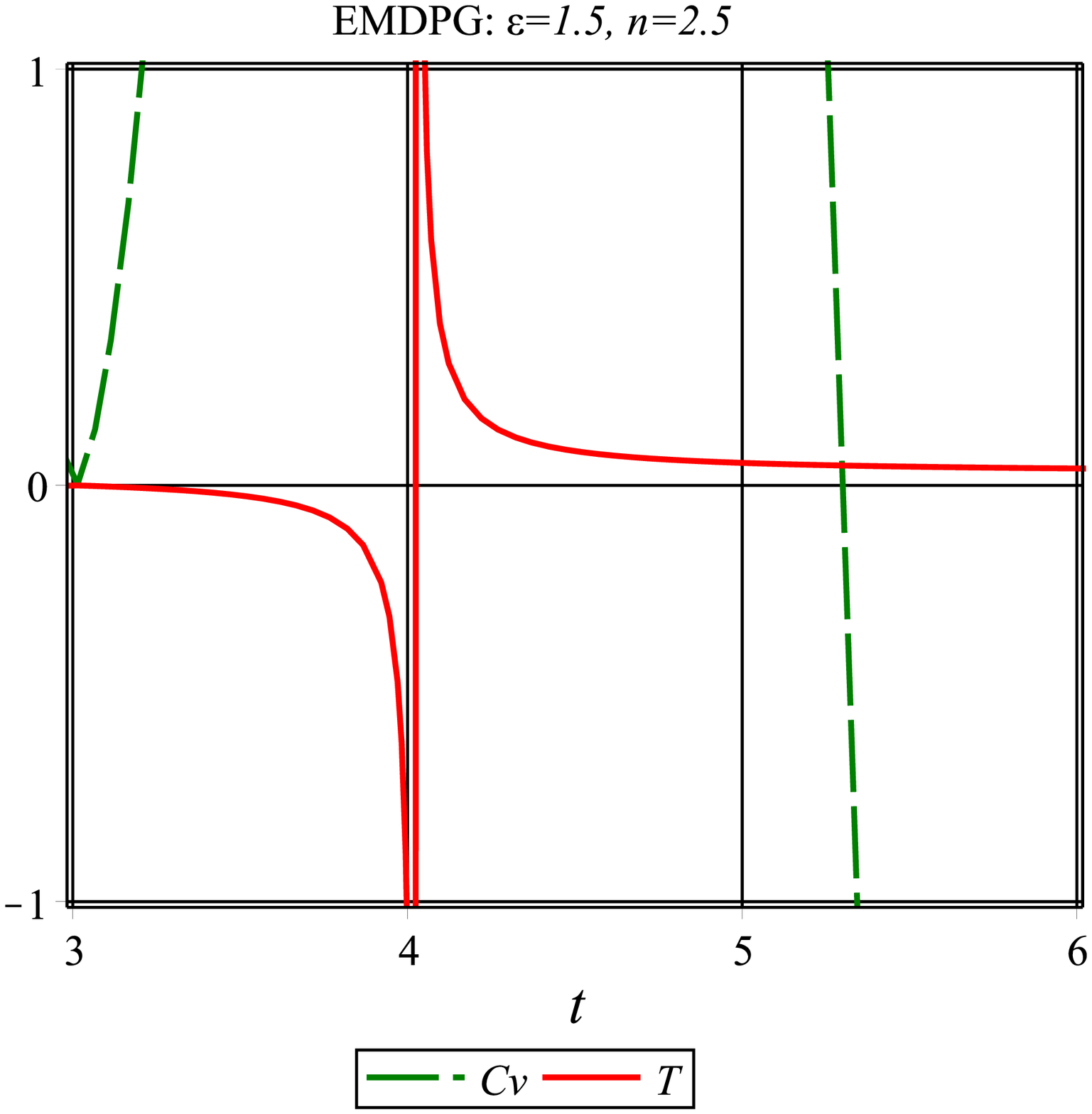}
 \end{array}$
 \end{center}
\caption{Specific heat of the EMDPG model in terms of time $t$ in
unit of $G$. The initial conditions are taken as $A=\alpha_{1}=1$,
and $\gamma=0.4$. (a) is for $0<\varepsilon<1$ from
(\ref{Conditions-EMDPG}); (b) is for $\varepsilon>1$ from
(\ref{Conditions-EMDPG-another}). Temperature has also been shown
in both the plots. The two other plots below show a clear picture
of Temperature $T$ in a zoomed range.}
 \label{fig1}
\end{figure}

For this model, using equation (\ref{submodel13}) we have,
\begin{equation}\label{fR-EMDPG}
f_{R}=n\alpha_{1}R^{n-1}
\end{equation}
and
\begin{equation}\label{fT-EMDPG}
f_{\mathbf{T^2}}=m\alpha_{2}({T^2})^{m-1}
\end{equation}
Using equation (\ref{fR-EMDPG}) in the relation (\ref{S}) one can
obtain the entropy of the system as,
\begin{equation}\label{S-EMDPG}
S=\frac{n\pi\alpha_{1}}{G}r_{A}^{2}\left[\frac{2|2-\dot{r}_{A}|}{r_{A}^{2}}\right]^{n-1}
\end{equation}
Then, using the equations (\ref{T}), (\ref{C}) and (\ref{S-EMDPG})
one can obtain the specific heat as,
\begin{equation}\label{C-EMDPG}
C_{V}=\frac{n\pi\alpha_{1}}{G}\left[\frac{2|2-\dot{r}_{A}|}{r_{A}^{2}}\right]^{n-1}\frac{(n-1)r_{A}{\ddot{r}}_{A}+2(n-2)\dot{r}_{A}|2-\dot{r}_{A}|}
{r_{A}{\ddot{r}}_{A}-{\dot{r}_{A}}^{2}+2\dot{r}_{A}}
\end{equation}
As we know that in order to have a stable model we should have
$C_{V}\geq0$. It will be the case if the following conditions are
satisfied simultaneously,
\begin{eqnarray}\label{Conditions-EMDPG}
(n-1)r_{A}{\ddot{r}}_{A}+2(n-2)\dot{r}_{A}|2-\dot{r}_{A}|\geq0\nonumber\\
r_{A}{\ddot{r}}_{A}-{\dot{r}_{A}}^{2}+2\dot{r}_{A}>0
\end{eqnarray}
From the second condition we get, ${r}_{A}< Ae^{\gamma
t}-\frac{2}{\gamma}$ where $A$ and $\gamma$ are arbitrary
constants. Without any loss of generality we can consider $A$ and
$\gamma$ as positive constants. Therefore, we introduce a small
constant $\varepsilon$ ($0<\varepsilon<1$) and choose the
following solution from the above obtained condition,
\begin{equation}\label{r-EMDPG}
{r}_{A}=\varepsilon(Ae^{\gamma t}-\frac{2}{\gamma})
\end{equation}
In that case, the denominator of (\ref{C-EMDPG}) is positive. But,
numerator of (\ref{C-EMDPG}) i.e., the first condition of
(\ref{Conditions-EMDPG}) yields,
\begin{eqnarray}\label{Conditions1-EMDPG}
(n-1)\varepsilon(A\gamma e^{\gamma t}-2)+2(n-2)|2-\varepsilon A\gamma e^{\gamma t}|\geq0.
\end{eqnarray}
This equation tells that relatively smaller values of $n$ ($n<2.5$
with our selected model parameters) yields partly stable model. In
fact, stability is exhibited at the late time and the model is
unstable initially. Initial instability may be attributed to the
particle creation process and phase transition era. It means that
the EMDPG model is initially in an unstable phase which transits
to the stable phase in late time. Obviously these time limits
depend on the values of the model parameters $\gamma$ and
$\varepsilon$, and by proper fine tuning we may alter these limits
as desired keeping an eye on the observational data. Finally, we
find that in our scale for the case of $n\geq 2.5$, the model is
completely stable at the late time where both $C_V>0$ and $T>0$.
These are illustrated in Fig. \ref{fig1}(a), where the typical
behavior of the specific heat at constant volume is represented.
Also, solid red line of Fig. \ref{fig1} show the temperature which
is independent of $n$. We see that beyond $t=6$ (late time), both
temperature and specific heat remain in the positive region.

There is also another possibility for the positivity of $C_V$ from
equation (\ref{C-EMDPG}) given as below,

\begin{eqnarray}\label{Conditions-EMDPG-another}
(n-1)r_{A}{\ddot{r}}_{A}+2(n-2)\dot{r}_{A}|2-\dot{r}_{A}|\leq0\nonumber\\
r_{A}{\ddot{r}}_{A}-{\dot{r}_{A}}^{2}+2\dot{r}_{A}<0
\end{eqnarray}
the result of which is represented in Fig.\ref{fig1}(b). It is
obtained if both the conditions of (\ref{Conditions-EMDPG}) be
negative. The result is again similar to (\ref{r-EMDPG}) with
$\varepsilon>1$ in this case. From Fig.\ref{fig1}(b) we can see
that, this yields an unstable model at the late time (present
epoch) where the specific heat shoots towards the negative region
for almost all values of $n$. So we are justified in ignoring such
a possibility. In order to get a clear idea of temperature we have
plotted it along with the specific heat in a zoomed range in the
other two plots of Fig. \ref{fig1}. In those figures we can
clearly see that the temperature remains in the positive level
thus indicating model stability.

Therefore, by using the equations (\ref{r}) and (\ref{r-EMDPG}) we
can obtain the cosmological scale factor as,
\begin{eqnarray}\label{a-1}
a(t)=a_{0}\left(A\gamma-2e^{-\gamma t}\right)^{\frac{1}{2\varepsilon}}.
\end{eqnarray}
Hence, by using the equation (\ref{q}) we can obtain,
\begin{eqnarray}\label{q-1}
q=-1+\varepsilon A\gamma e^{\gamma t}.
\end{eqnarray}
In order to continue our thermodynamic analysis we need to include
the expressions for energy density $\rho_{eff}$ and pressure
$p_{eff}$ respectively from (\ref{rho}) and (\ref{p}). We can see
from section 2.1 that we have two different cases of $w=-1$ and
$w=-\frac{1}{3}$ where we may find values for energy densities and
pressure for the EMSG model. We will study the two cases
separately.

\subsubsection{$\Lambda CDM$}

Using the relation (\ref{fT-EMDPG}) in the first solution given by
(\ref{cons1}) confirms the second one which is a constant as
follow,
\begin{equation}\label{C01}
\rho=\left(\frac{m\alpha_{2}}{4^{1-m}}\right)^{\frac{1}{1-2m}}\equiv
C_{01}\equiv C_{0}
\end{equation}
and hence,
\begin{equation}
p=-C_{01}\equiv -C_{0}
\end{equation}
Our numerical analysis show that the first law of thermodynamics
for this model is satisfied at the initial times if we choose
small values for $A$ and $C_{01}$. In the case of $A=0$ the first
law of thermodynamics is satisfied completely. We find that
$w_{eff}$ yields a negative unity at the late time corresponding
to $\Lambda CDM$ era. By using the equation (\ref{F}) we can
obtain Helmholtz free energy as following,
\begin{equation}
F=-\frac{An\alpha_{1}\gamma\varepsilon(A\gamma e^{\gamma t}-2)\varrho_{1}^{n}}{32\pi G}-\frac{4\pi\gamma^{3}\varepsilon^{3}(A\gamma e^{\gamma t}-2)^{3}\varrho_{11}}{3n\alpha_{1}\varrho_{1}^{n-1}},
\end{equation}
where $\varrho_{1}$ and $\varrho_{11}$ defined in the appendix
section. In the Fig. \ref{fig2} we can see typical behavior of the
Helmholtz free energy for the EMDPG model with $w=-1$. We find
that the cases of $m=0.5$ and $m=2$ have similar behavior. We can
see a local minimum at the late time which may correspond to the
model stability at the late time. However, in this epoch the first
law of thermodynamics is not satisfied and hence there is a
confusion regarding stability in this phase. So we are motivated
to explore other models, but before that we consider briefly the
case with $w=-\frac{1}{3}$ for this model.

\begin{figure}[h!]
 \begin{center}$
 \begin{array}{cccc}
\includegraphics[width=55 mm]{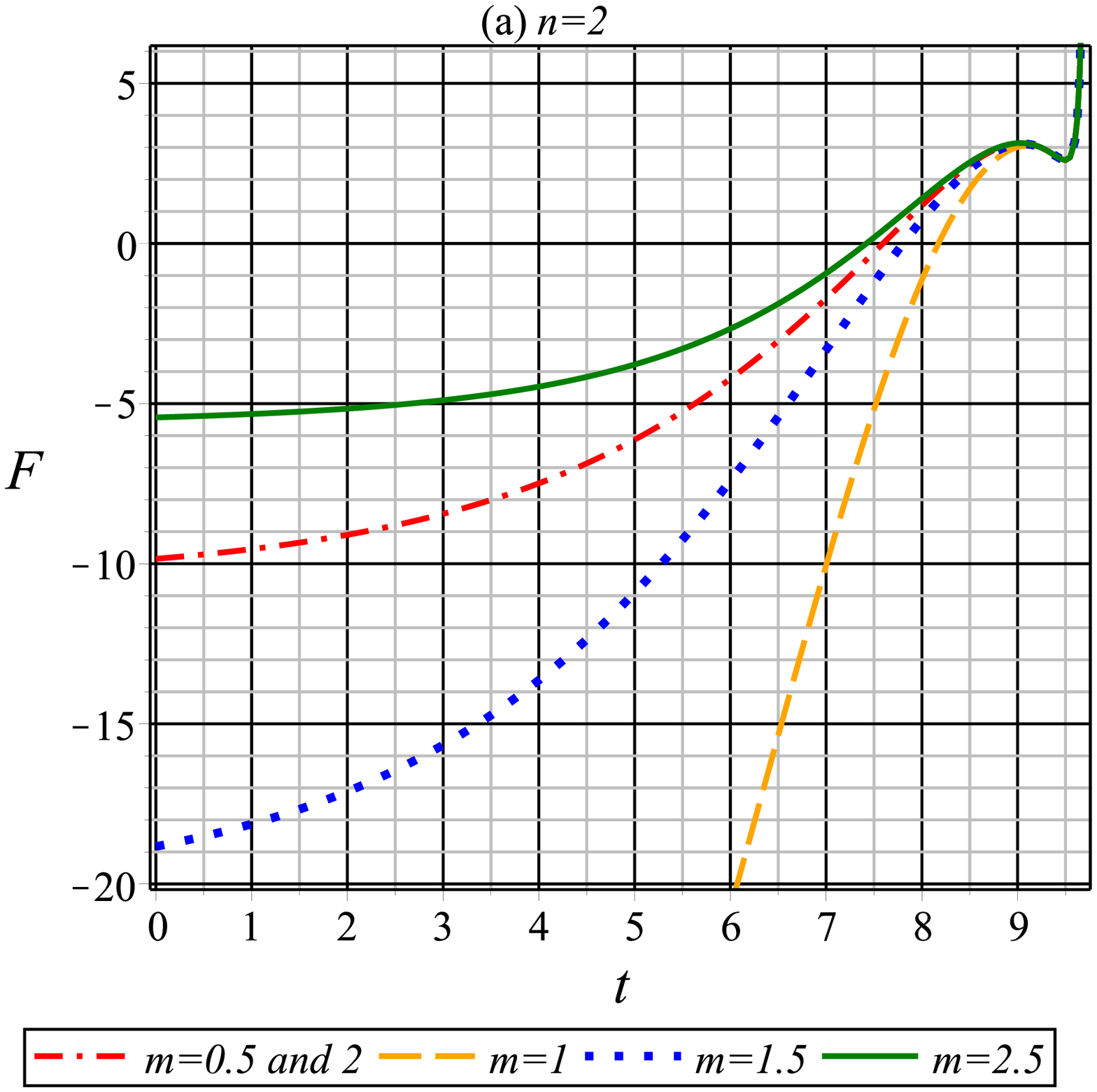}\includegraphics[width=55 mm]{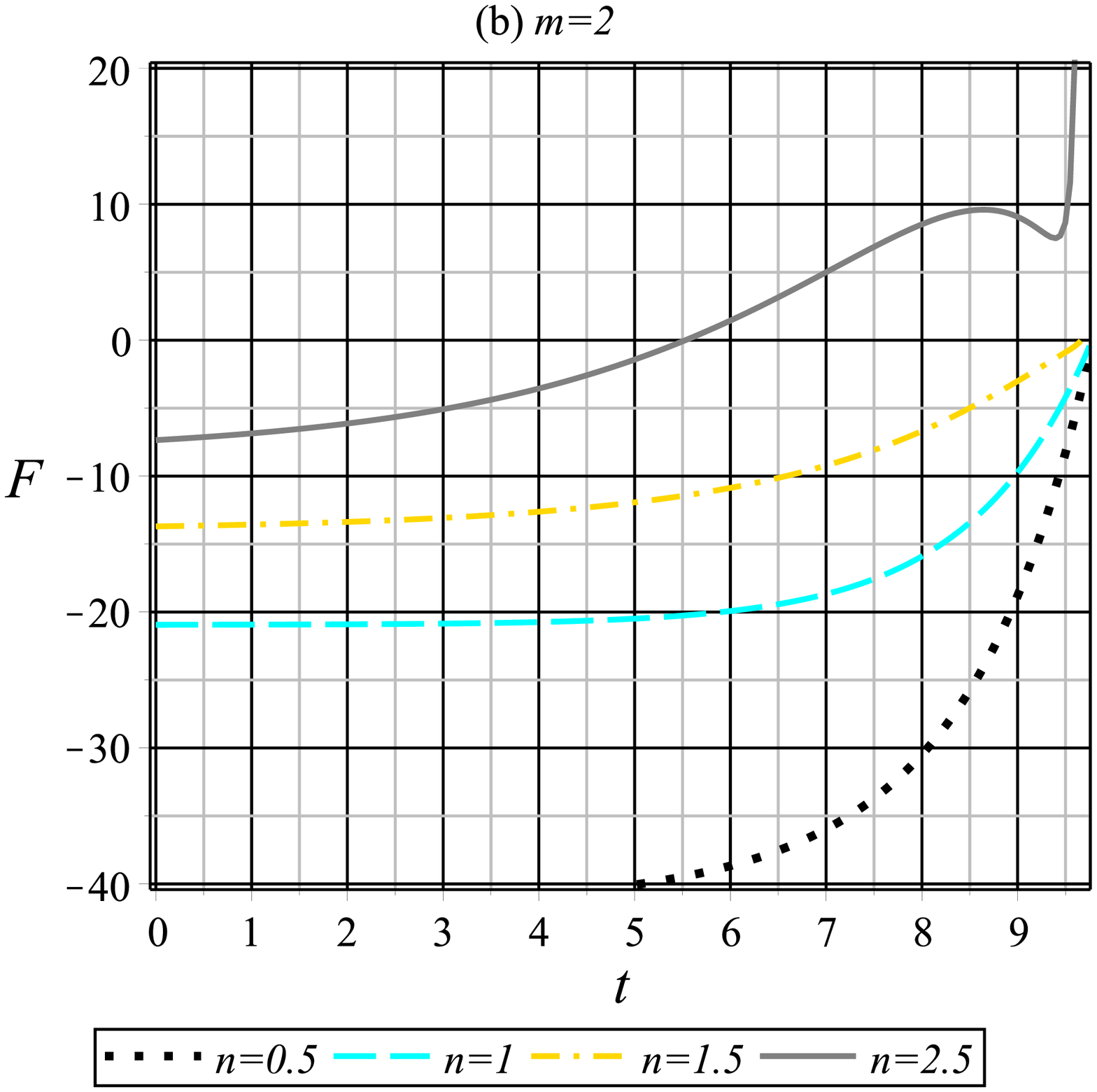}
 \end{array}$
 \end{center}
\caption{Helmholtz free energy of the EMDPG model with respect to
time $t$ for with $w=-1$ in unit of $G$. The initial conditions
are taken as $\alpha_{1}=\alpha_{2}=1$, $A=0.1$, $\gamma=0.4$ and
$\varepsilon=0.5$.}
 \label{fig2}
\end{figure}

\subsubsection{$w=-\frac{1}{3}$}

Using the relation (\ref{fT-EMDPG}) in the solution given by (\ref{cons2}) one can obtain,
\begin{equation}
a=\left[\rho^{3}e^{3m\alpha_{2}(\frac{4}{3})^{m}\rho^{2m-1}-C_{1}}\right]^{-\frac{1}{6}}
\end{equation}
or
\begin{equation}
z=\left[\rho^{3}e^{3m\alpha_{2}(\frac{4}{3})^{m}\rho^{2m-1}-C_{1}}\right]^{\frac{1}{6}}-1
\end{equation}
It yields the following equation,
\begin{equation}
H=\frac{1}{r_{A}}=\frac{\left(3m\alpha_{2}(\frac{4}{3})^{m}(m-\frac{1}{2})\rho^{2m-1}-\frac{3}{2}\right)\dot{\rho}}{3\rho}
\end{equation}
Hence, we have the cosmic apparent horizon radius in terms of $\rho$ and
its derivative. In order to have analytical solutions we consider
special case of $m=\frac{1}{2}$, (which was physically identical
to the case of $m=2$ in the previous case), following which we can
write,
\begin{equation}
\rho=\frac{\rho_{0}}{a^{2}}
\end{equation}
where
\begin{equation}
\rho_{0}=\left[e^{\frac{3}{2}\alpha_{2}\sqrt{\frac{4}{3}}-C_{1}}\right]^{-\frac{1}{3}},
\end{equation}
is a constant. It is clear that increasing time, increases scale factor and decreases density.
Then, using (\ref{a-1}) we can obtain time dependent energy density.\\
By using the equation (\ref{F}) we can obtain Helmholtz free
energy as,
\begin{equation}
F=-\frac{An\alpha_{1}\gamma\varepsilon(A\gamma e^{\gamma t}-2)\varrho_{1}^{n}}{32\pi G}-\frac{4\pi\gamma^{3}\varepsilon^{3}(A\gamma e^{\gamma t}-2)^{3}\varrho_{12}}{3n\alpha_{1}\varrho_{1}^{n-1}},
\end{equation}
where $\varrho_{1}$ and $\varrho_{12}$ are defined in appendix. In
the Fig.\ref{fig3} we have obtained the typical behavior of the
Helmholtz free energy and see the occurrence of a minimum which
may be a sign of model stability.

\begin{figure}[h!]
 \begin{center}$
 \begin{array}{cccc}
\includegraphics[width=55 mm]{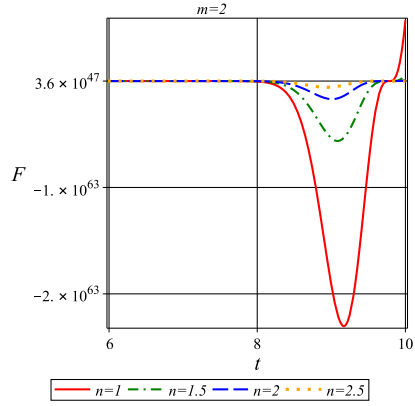}
 \end{array}$
 \end{center}
\caption{Typical behavior of the Helmholtz free energy for the
EMDPG model with $w=-\frac{1}{3}$ in unit of $G$ against time $t$.
The parameters are considered as $m=2$, $\alpha_{1}=\alpha_{2}=1$,
$A=0.1$, $\gamma=0.4$, $\varepsilon=0.5$ and unit value for other
constant.}
 \label{fig3}
\end{figure}

\begin{figure}[h!]
 \begin{center}$
 \begin{array}{cccc}
\includegraphics[width=70 mm]{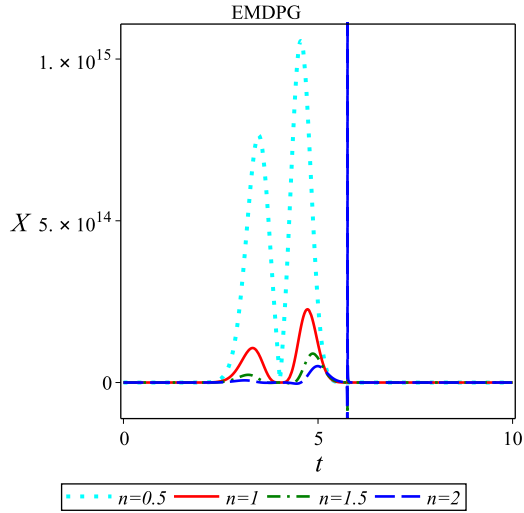}
 \end{array}$
 \end{center}
\caption{The first law of thermodynamics of the EMDPG model with
$w=-\frac{1}{3}$ in unit of $G$ versus time. The initial
conditions are $m=2$, $\alpha_{1}=\alpha_{2}=1$, $A=0.1$,
$\gamma=0.4$, $\varepsilon=0.5$ and unit values for other
constants.}
 \label{fig4}
\end{figure}

We find that this model satisfies the first law of thermodynamics
at both the early and late times. But there is an intermediate era
where the first law of thermodynamics is violated for a short
period of time. It may correspond to the time of structure
formation. In order to demonstrate it by numerical analysis, we
rewrite (\ref{WT}) as follows,
\begin{eqnarray}\label{W2}
X\equiv\bar{T}dS-dU+WdV.
\end{eqnarray}
Then we see the evolution of $X$ versus time $t$ as represented in
Fig.\ref{fig4}. From our construction of equation (\ref{W2}) we see
that when $X=0$ then the first law of thermodynamics is satisfied,
otherwise it is violated. From the plot we can see that there is
an intermediate phase of violation of the first law as stated
above. In these thermodynamic studies of the EMSG models we indeed
need effective energy density and effective pressure which we have
provided in appendix for the reader's convenience.

\subsection{Thermodynamics in EMDEG model}

\begin{figure}[h!]
 \begin{center}$
 \begin{array}{cccc}
\includegraphics[width=70 mm]{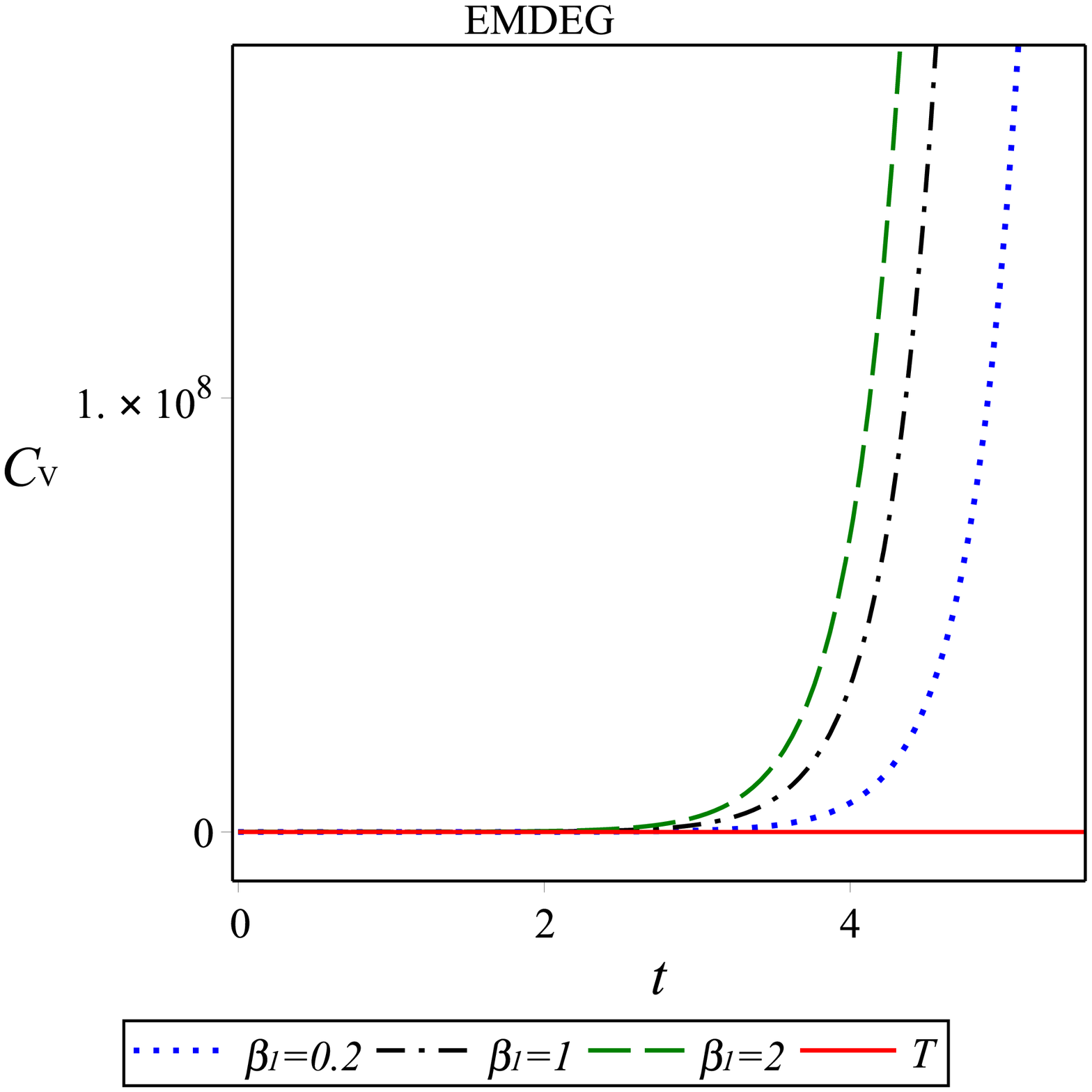}~~~~~~\includegraphics[width=70 mm]{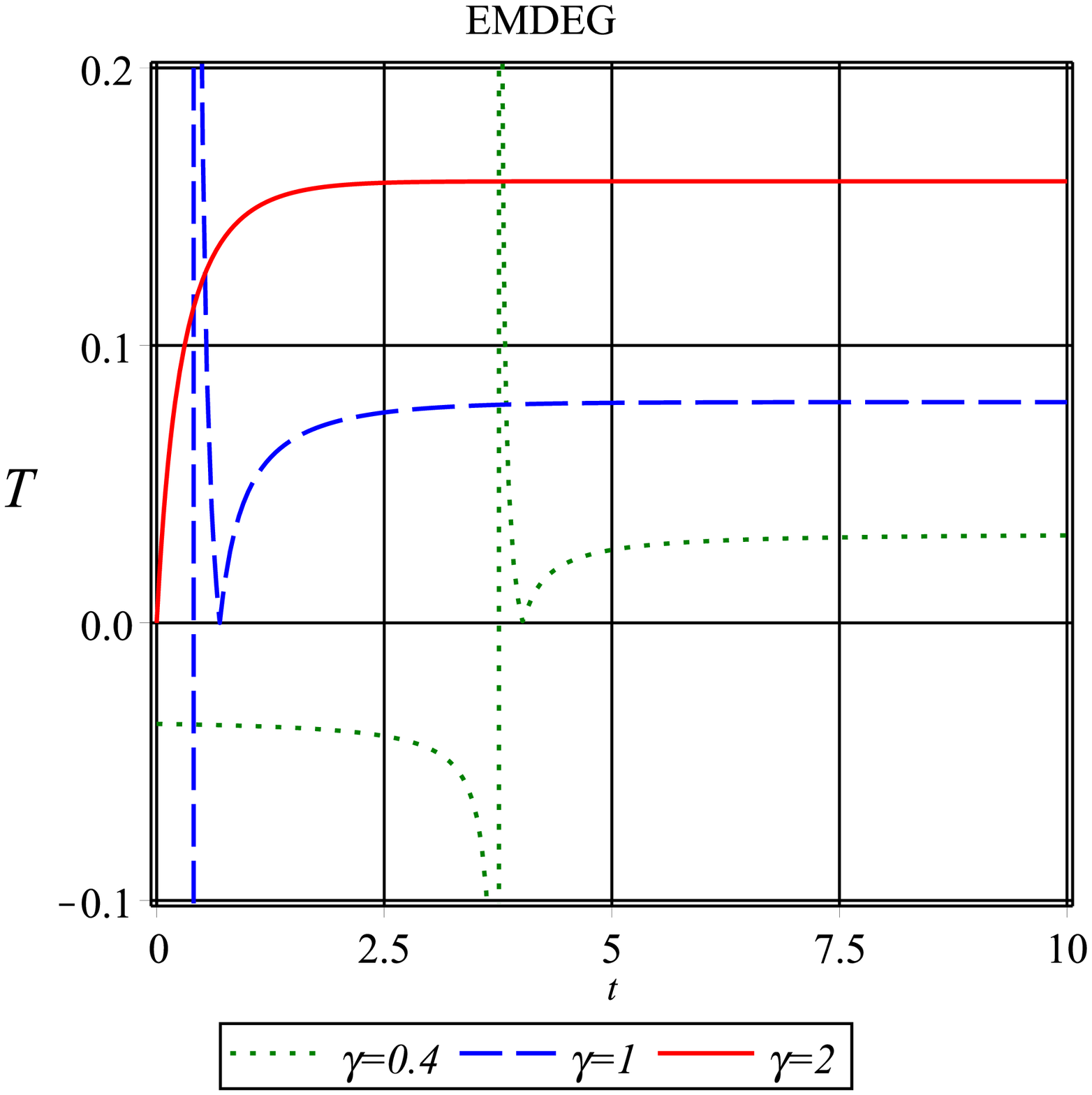}
 \end{array}$
 \end{center}
\caption{Specific heat of the EMDEG model in unit of $G$ versus
time for $\gamma=g_{1}=1$. Temperature has also been shown in the
plot. The plot on the right shows temperature in the zoomed
range.}
 \label{fig5}
\end{figure}

In this model, by using the equation (\ref{submodel14}) we have,
\begin{equation}\label{fR-EMDEG}
f_{R}=g_{1}\beta_{1}e^{\beta_{1}R}
\end{equation}
and
\begin{equation}\label{fT-EMDEG}
f_{\mathbf{T^2}}=g_{2}\beta_{2}e^{\beta_{2}T^{2}}
\end{equation}
Using the equations (\ref{fR-EMDEG}) and (\ref{Ricci3}) in the
relation (\ref{S}) one can the obtain entropy as follows,
\begin{equation}\label{S-EMDEG}
S=\frac{\pi
g_{1}\beta_{1}}{G}r_{A}^{2}\exp\left(\beta_{1}\left[\frac{2|2-\dot{r}_{A}|}{r_{A}^{2}}\right]\right)
\end{equation}
Then, using the equations (\ref{T}), (\ref{C}) and (\ref{S-EMDEG})
one can obtain the specific heat as,
\begin{equation}\label{C-EMDEG}
C_{V}=\frac{2\pi
g_{1}\beta_{1}}{G}|2-\dot{r}_{A}|\frac{\beta_{1}r_{A}{\ddot{r}}_{A}+4\beta_{1}\dot{r}_{A}-2\beta_{1}{\dot{r}_{A}}^{2}-\dot{r}_{A}
r_{A}^{2}}
{r_{A}{\ddot{r}}_{A}-{\dot{r}_{A}}^{2}+2\dot{r}_{A}}\exp\left(\beta_{1}\left[\frac{2|2-\dot{r}_{A}|}{r_{A}^{2}}\right]\right)
\end{equation}
As before, we should have $C_{V}\geq0$ for stability. It will be
realized if the following conditions are satisfied simultaneously,
\begin{eqnarray}\label{Conditions-EMDEG}
\beta_{1}r_{A}{\ddot{r}}_{A}+4\beta_{1}\dot{r}_{A}-2\beta_{1}{\dot{r}_{A}}^{2}-\dot{r}_{A} r_{A}^{2}\geq0\nonumber\\
r_{A}{\ddot{r}}_{A}-{\dot{r}_{A}}^{2}+2\dot{r}_{A}\geq0
\end{eqnarray}
Although, we can also assume both equations as negative valued
which yields to the similar result as the previous model with a
change in the range of the parameter $\epsilon$. We can see that
the second condition is the same as previous case and satisfied
with the same solution ${r}_{A}\leq Ae^{\gamma
t}-\frac{2}{\gamma}$, where as before $A$ and $\gamma$ are some
constants. Therefore, we can consider $A$ and $\gamma$ as positive
constants without any loss of generality. However the first
condition is different from the previous case and we find that
both conditions are satisfied with the following solution,
\begin{equation}\label{r-EMDEG}
{r}_{A}=Ae^{\gamma t}-\frac{2}{\gamma}+\epsilon
\end{equation}
where $\epsilon$ is a positive constant. Using the constraint of
temperature being a positive quantity, we find the lower bound of
this parameter as $\epsilon\geq\frac{2}{\gamma}$. Without the loss
of generality we can choose $\epsilon=\frac{2}{\gamma}$ to find,
\begin{equation}\label{r-EMDEG2}
{r}_{A}=Ae^{\gamma t}
\end{equation}
Using this we find,
\begin{equation}\label{T-EMDEG2}
T=\frac{|2-\gamma{r}_{A}|}{4\pi {r}_{A}}
\end{equation}
which is a positive quantity. In that case the first condition of
(\ref{Conditions-EMDEG}) is reduced to the following equation,
\begin{equation}\label{Conditions-EMDEG1}
\beta_{1}\gamma{r}_{A} (\gamma{r}_{A}-4)+\gamma{r}_{A}^{3}\geq0
\end{equation}
It satisfied for infinitesimal $\beta_{1}$ or $\gamma A\geq4$.
Hence, we choose $A=\frac{4}{\gamma}$ and consider the following
solution from equation (\ref{r-EMDEG2}),

\begin{equation}\label{r-EMDEG3}
{r}_{A}=\frac{4}{\gamma}e^{\gamma t}
\end{equation}
In that case the specific heat is a completely positive quantity
as follows,
\begin{equation}\label{C-EMDEG2}
C_{V}=\frac{8\pi g_{1}\beta_{1}|1-2e^{\gamma
t}|(\beta_{1}\gamma^{2}(e^{\gamma t}-1)+4e^{2\gamma
t})e^{\frac{\beta_{1}\gamma^{2}|1-2e^{\gamma t}|}{4e^{2\gamma
t}}}}{\gamma^{2}G}
\end{equation}
In the Fig. \ref{fig5} we can see the typical behavior of specific
heat which is completely positive, indicating that this model is
more stable than the previous one. The plot on the right shows the
temperature in the zoomed range, where we see that it remains
perfectly in the positive region showing model stability.

Therefore, by using the equations (\ref{r}) and (\ref{r-EMDEG3}) we can obtain scale factor as,
\begin{eqnarray}\label{a-2}
a(t)=a_{0}e^{-\frac{1}{4}e^{-\gamma t}}
\end{eqnarray}
Then, by using the equation (\ref{q}) we can obtain the
deceleration parameter as,
\begin{eqnarray}\label{q-2}
q=-1+4 e^{\gamma t}
\end{eqnarray}
Now we use the energy densities and the pressure of matter for the
two cases separately.

\subsubsection{$\Lambda CDM$}

Using the relation (\ref{fT-EMDEG}) in the first solution given by
(\ref{cons1}), confirms the second one as given below
\begin{equation}\label{C02}
\rho=\frac{1}{g_{2}\beta_{2}e^{\frac{1}{2}W[\frac{8}{\beta_{2}g_{2}^{2}}]}}\equiv
C_{02} \equiv C_{0}
\end{equation}
where $W[y]$ is the Lambert W function, and hence we have,
\begin{equation}
p=-C_{02}\equiv C_{0}
\end{equation}
where $C_{02}$ denotes a constant of the second model. Similar to
the previous case, we find that the first law of thermodynamics is
satisfied initially while being violated at the late time. We find
that larger values of $\gamma$ yields a more stable period. By
using the equation (\ref{F}) we can obtain Helmholtz free energy
as following,
\begin{equation}
F=-\frac{2\beta_{1}e^{\gamma t}}{\gamma}|2e^{\gamma
t}-1|\varrho_{2}-\frac{256\pi e^{3\gamma
t}}{3g_{1}\beta_{1}\gamma^{3}\varrho_{2}}\varrho_{21}
\end{equation}
where $\varrho_{2}$ and $\varrho_{21}$ are defined in appendix.
Helmholtz free energy of this model is an increasing function of
time.

\subsubsection{$w=-\frac{1}{3}$}

\begin{figure}[h!]
 \begin{center}$
 \begin{array}{cccc}
\includegraphics[width=55 mm]{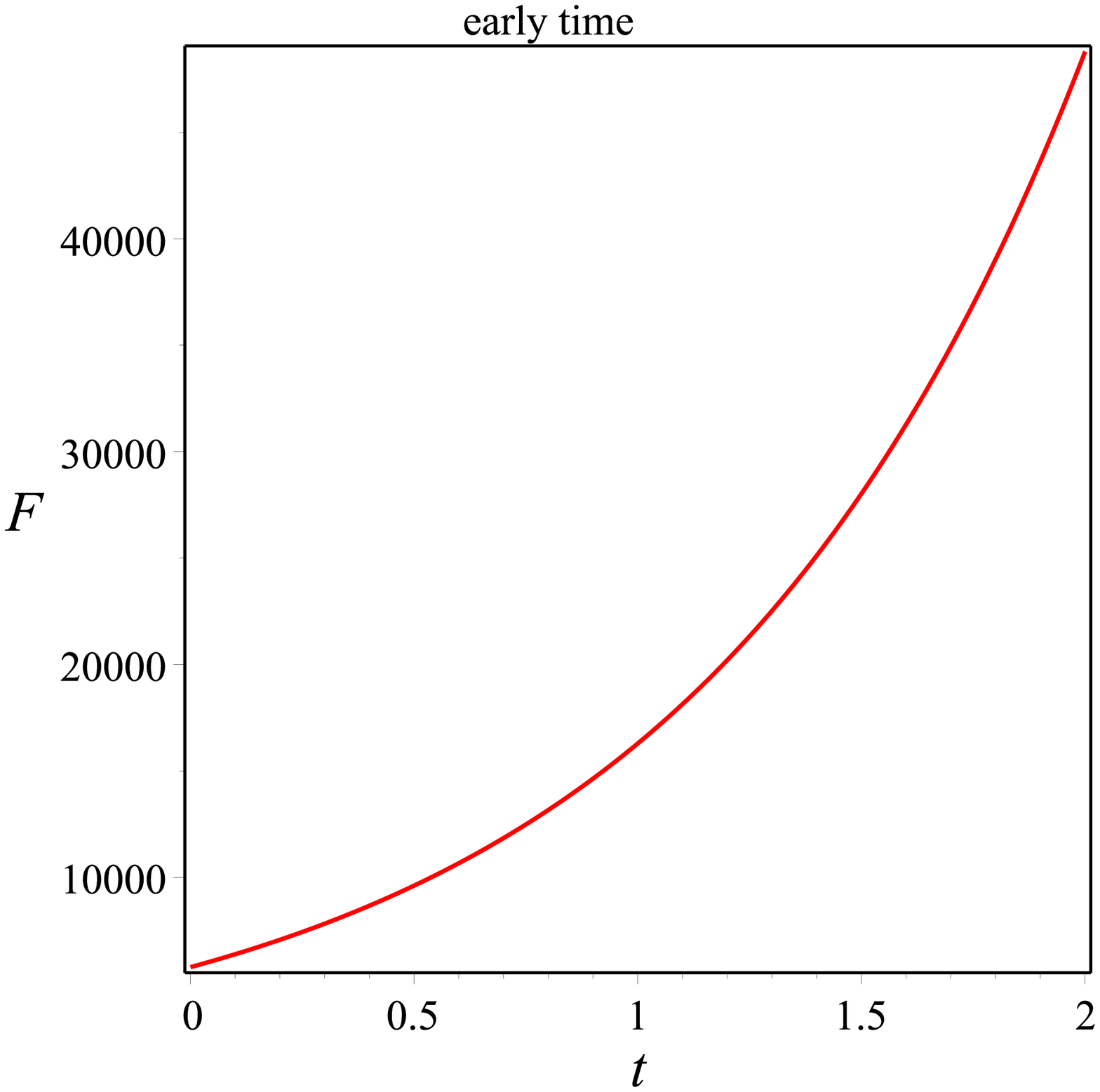}\includegraphics[width=55 mm]{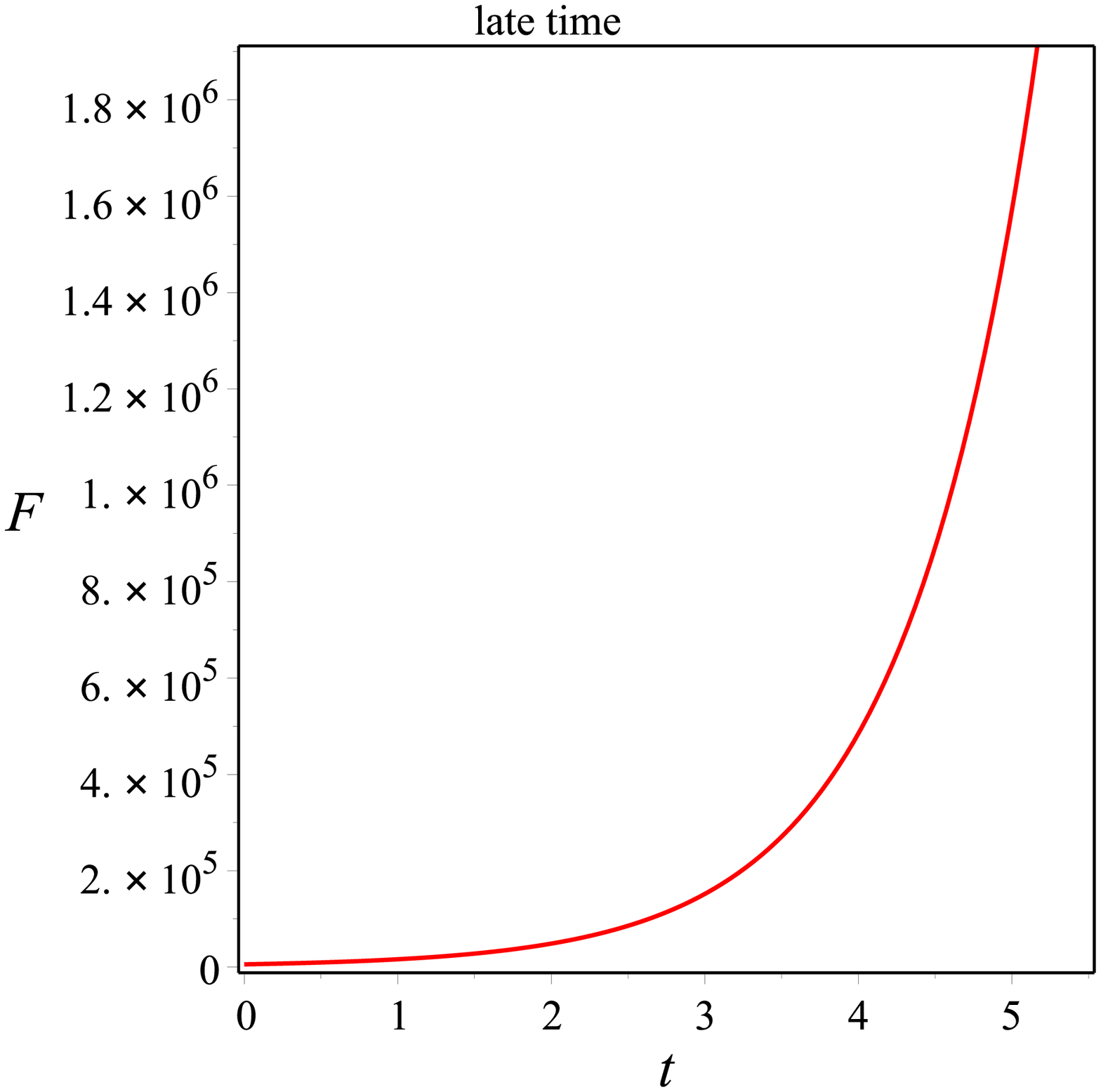}
 \end{array}$
 \end{center}
\caption{Typical behavior of the Helmholtz free energy of the
EMDEG model with $w=-\frac{1}{3}$ in unit of $G$ versus time for
$g_{1}=g_{2}=1$, $\beta_{1}=\beta_{2}=1$, $\gamma=0.4$ and unit
value for other constants.}
 \label{fig6}
\end{figure}

Using the relation (\ref{fT-EMDEG}) in the solution given by (\ref{cons2}), one can obtain scale factor in terms of energy density as,
\begin{equation}
a=\left[\rho^{3}\left(e^{-\frac{4}{3}\rho
g_{2}\beta_{2}e^{\frac{4}{3}\rho^{2}}}\right)^{3}e^{C_{2}}\right]^{-\frac{1}{6}}
\end{equation}
which is used to obtain the following redshift,
\begin{equation}
z=\left[\rho^{3}e^{-4\rho
g_{2}\beta_{2}e^{\frac{4}{3}\rho^{2}}+C_{1}}\right]^{\frac{1}{6}}-1
\end{equation}
where $C_{1}$ and $C_{2}$ are some integration constants. It
yields the following equation for the Hubble expansion parameter,
\begin{equation}
H=\frac{1}{r_{A}}=\frac{16\left(g_{2}\beta_{2}\rho(\frac{3}{8}+\rho^{2})e^{\frac{4}{3}\rho^{2}}-\frac{9}{32}\right)\dot{\rho}}{9\rho}
\end{equation}
Hence, we have the cosmic apparent horizon in terms of $\rho$ and its derivative. In order to have an analytical relation we consider special cases of the early and the late times.\\
At the early time, we assume $\rho\gg1$ and find,
\begin{equation}
\rho\approx\frac{3(\gamma_{2}-e^{-\gamma
t})}{8g_{2}\beta_{2}e^{\frac{1}{2}W[-\frac{3(\gamma_{2}-e^{-\gamma
t})}{8\beta_{2}^{2}g_{2}^{2}}]}}
\end{equation}
where $W[y]$ is the Lambert W function, and
$\gamma_{2}=C_{2}\gamma$ with $C_{2}$ is an integration constant.
It yields the following Helmholtz free energy,
\begin{equation}
F=-\frac{2\beta_{1}e^{\gamma t}}{\gamma}|2e^{\gamma
t}-1|\varrho_{2}-\frac{256\pi e^{3\gamma
t}}{3g_{1}\beta_{1}\gamma^{3}\varrho_{2}}\varrho_{22}
\end{equation}
and see appendix for definition of $\varrho_{2}$ and $\varrho_{22}$.
On the other hand, at the late time we assume $\rho\ll1$ and find,
\begin{equation}
\rho\approx C_{2}e^{\frac{1}{2}e^{-\gamma t}}
\end{equation}
It yields the following expression for the Helmholtz free energy,
\begin{equation}
F=-\frac{2\beta_{1}e^{\gamma t}}{\gamma}|2e^{\gamma
t}-1|\varrho_{2}-\frac{256\pi e^{3\gamma
t}}{3g_{1}\beta_{1}\gamma^{3}\varrho_{2}}\varrho_{23}
\end{equation}
and see appendix for definition of $\varrho_{2}$ and
$\varrho_{23}$. In the plots of the Fig. \ref{fig6} we can see
typical behavior of Helmholtz free energy at the early and the
late time. We can see that Helmholtz free energy is increasing
function of time. Regarding the first law of thermodynamics we
find similar result with the previous model. It means that the
first law of thermodynamics is satisfied both at the late and the
early times. Analyzing the entropy we see that the second law of
thermodynamics is satisfied too which means that the model entropy
(\ref{S-EMDEG}) is an increasing function of time.

\subsection{Thermodynamics in EMTPG model}

In this model, by using the equation (\ref{submodel21}) we have,
\begin{equation}\label{fR-EMTPG}
f_{R}=n\alpha_{1}R^{n-1}+m\alpha_{2}R^{m-1}(T^{2})^{l},
\end{equation}
and
\begin{equation}\label{fT-EMTPG}
f_{\mathbf{T^2}}=l\alpha_{2}R^{m}(T^{2})^{l-1}
\end{equation}
Hence, if we use the equation (\ref{fR-EMTPG}) in the entropy
expression (\ref{S}) we get,
\begin{equation}\label{S-EMTPG}
S=\frac{\pi}{G}r_{A}^{2}\left(n\alpha_{1}\left[\frac{2|2-\dot{r}_{A}|}{r_{A}^{2}}\right]^{n-1}
+m\alpha_{2}\left[\frac{2|2-\dot{r}_{A}|}{r_{A}^{2}}\right]^{m-1}((1+3w^{2})\rho^{2})^{l}\right).
\end{equation}
We can see that, unlike the previous models, here the entropy is
dependant on the energy density and hence we need explicit form of
the energy density, which is dependant on the cosmological era
(value of $w$). So we proceed to study the cases as before.

\subsubsection{$\Lambda CDM$}

\begin{figure}[h!]
 \begin{center}$
 \begin{array}{cccc}
\includegraphics[width=70 mm]{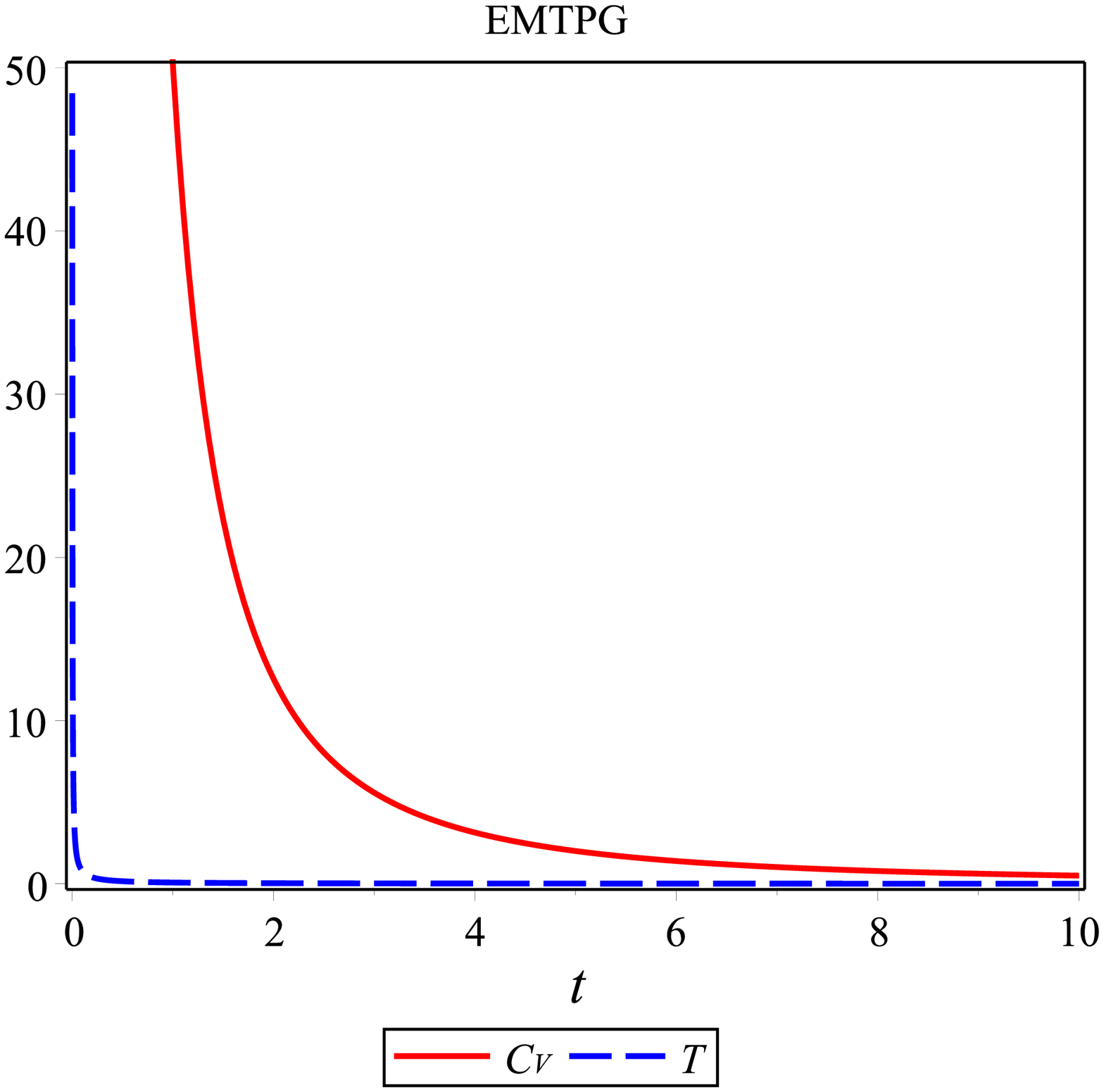}~~~~~\includegraphics[width=70 mm]{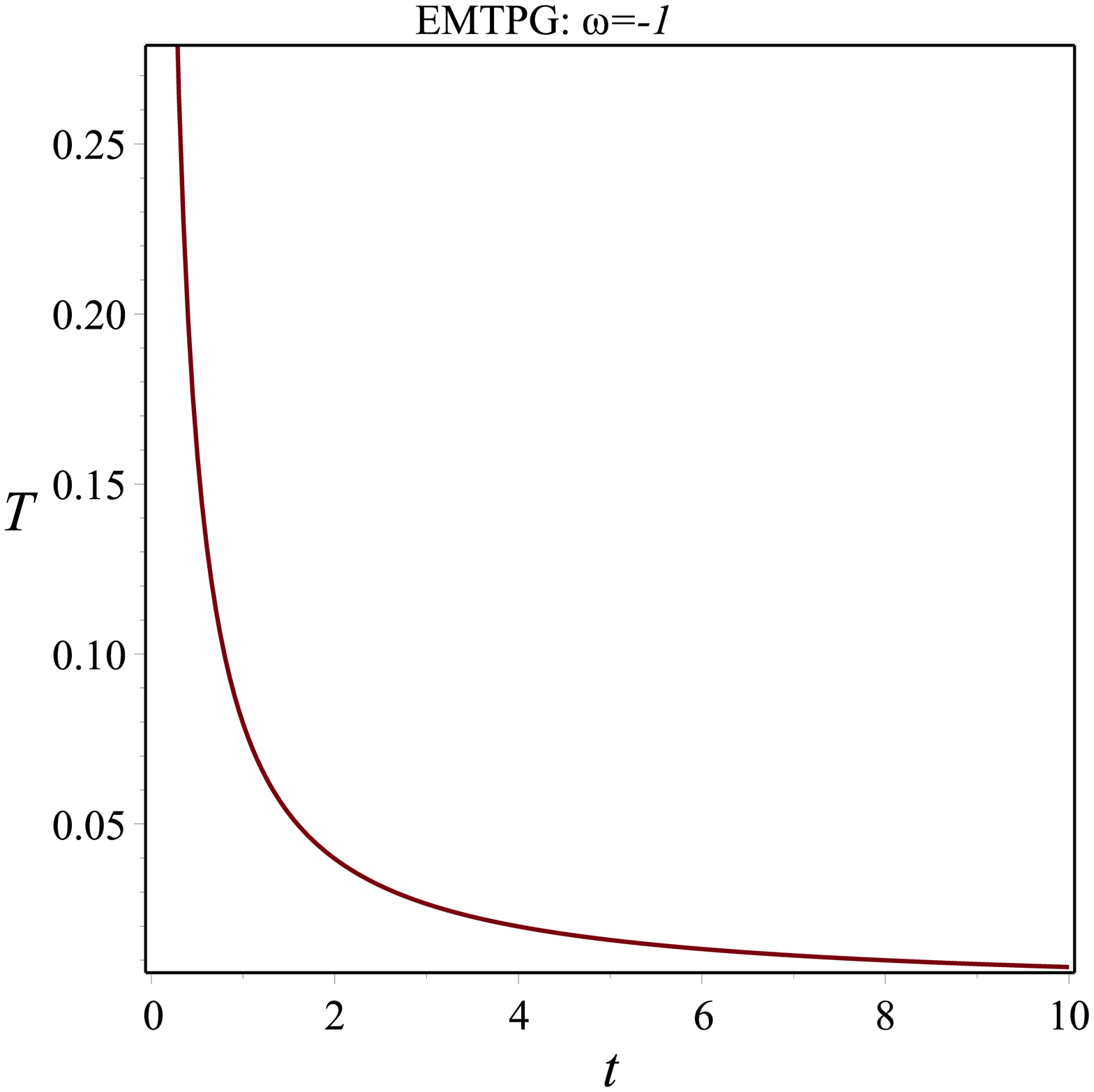}
 \end{array}$
 \end{center}
\caption{Specific heat of the EMTPG model with $w=-1$ in unit of
$G$ versus time for $M=N=2$ and unit value of other parameters.
Temperature has also been plotted in the figure. The figure on the
right shows temperature in the zoomed range.}
 \label{fig7}
\end{figure}

In this case, by using the first solution of (\ref{cons1}) one can
obtain,
\begin{equation}\label{dens-EMTPG1}
\rho=\left(\frac{1}{4^{l-1}l\alpha_{2}
R^{m}}\right)^{\frac{1}{2l-1}}
\end{equation}
Therefore, using the equations (\ref{dens-EMTPG1}) and (\ref{Ricci3}) in the equation (\ref{S-EMTPG}) we can write,
\begin{equation}\label{S-EMTPG1}
S=\frac{\pi}{G}r_{A}^{2}\left(n\alpha_{1}\left[\frac{2|2-\dot{r}_{A}|}{r_{A}^{2}}\right]^{n-1}
+m\alpha_{2}\left[\frac{2|2-\dot{r}_{A}|}{r_{A}^{2}}\right]^{\frac{2l(m-1)-3m+1}{2l-1}}4^{l}\left(\frac{1}{4^{l-1}l\alpha_{2}}\right)^{\frac{2l}{2l-1}}\right)
\end{equation}
It can be simplified as,
\begin{equation}\label{S-EMTPG1-1}
S=r_{A}^{2}\left(N_{0}R^N+M_{0}R^{M}\right)
\end{equation}
where $N=n-1$,~ $M=\frac{2l(m-1)-3m+1}{2l-1}$,~
$N_{0}=\frac{\pi}{G}n\alpha_{1}$ ~ and~
$M_{0}=\frac{\pi}{G}m\alpha_{2}4^{l}\left(\frac{1}{4^{l-1}l\alpha_{2}}\right)^{\frac{2l}{2l-1}}$
are constants. In order to satisfy the second law of
thermodynamics we should have
\begin{equation}\label{Second-law-EMTPG1}
\frac{dS}{dt}\geq0
\end{equation}
We see that the equation (\ref{Second-law-EMTPG1}) is satisfied if
we choose,
\begin{equation}\label{r-3}
r_{A}=r_{01}t+\frac{r_{02}}{t}
\end{equation}
where $r_{01}$ and $r_{02}$ are arbitrary constants. In this case,
suitable values of $M$ and $N$ can yield a stable model. For
example, in the Fig. \ref{fig7}, we can see typical behavior of
the specific heat for $M=N=2$. We find that larger values of these
parameters also yield positive specific heat. Also, from dashed
blue line of Fig. \ref{fig7} we can see that temperature is
positive in this model. To get a better idea about this we have
plotted temperature in a zoomed range in the figure on the right.
From the figure it is clearly evident that temperature is positive
showing the stability of the model. The entropy at the early time
($t\ll1$) may be written in the following form,
\begin{equation}\label{S-3}
S\approx \frac{X_{0}}{t^{2}}
\end{equation}
where
\begin{equation}\label{X0}
X_{0}=N_{0}e^{N\ln{2}}+M_{0}e^{M\ln{2}}
\end{equation}
is a constant. Moreover the specific heat of the early time
($t\ll1$) may be written in the following form,
\begin{equation}\label{C-3}
C_{V}\approx X_{0}+\frac{X_{1}}{t^{2}}
\end{equation}
where the constant $X_{0}$ is given by the equation (\ref{X0}) and
$X_{1}$ is a constant depending on the model parameters. In this
model the scale factor is obtained as,
\begin{equation}\label{a-3}
a=a_{0}(r_{01}t^{2}+r_{02})^{\frac{1}{2r_{01}}}
\end{equation}
Then, the deceleration parameter may be given by,
\begin{equation}\label{q-3}
q=-1+r_{01}-\frac{r_{02}}{t^{2}}
\end{equation}

\subsubsection{$w=-\frac{1}{3}$}

\begin{figure}[h!]
 \begin{center}$
 \begin{array}{cccc}
\includegraphics[width=60 mm]{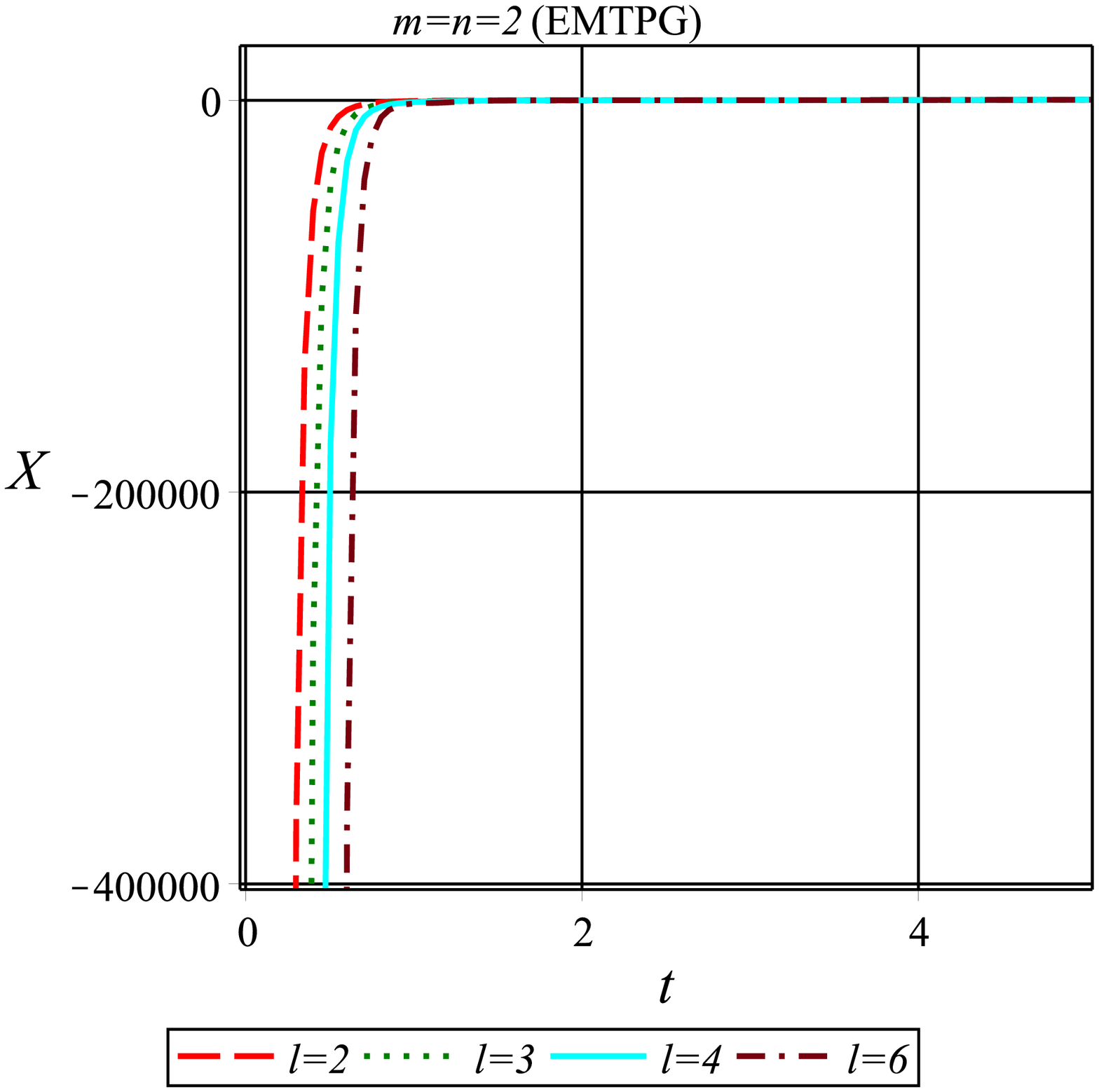}\includegraphics[width=60 mm]{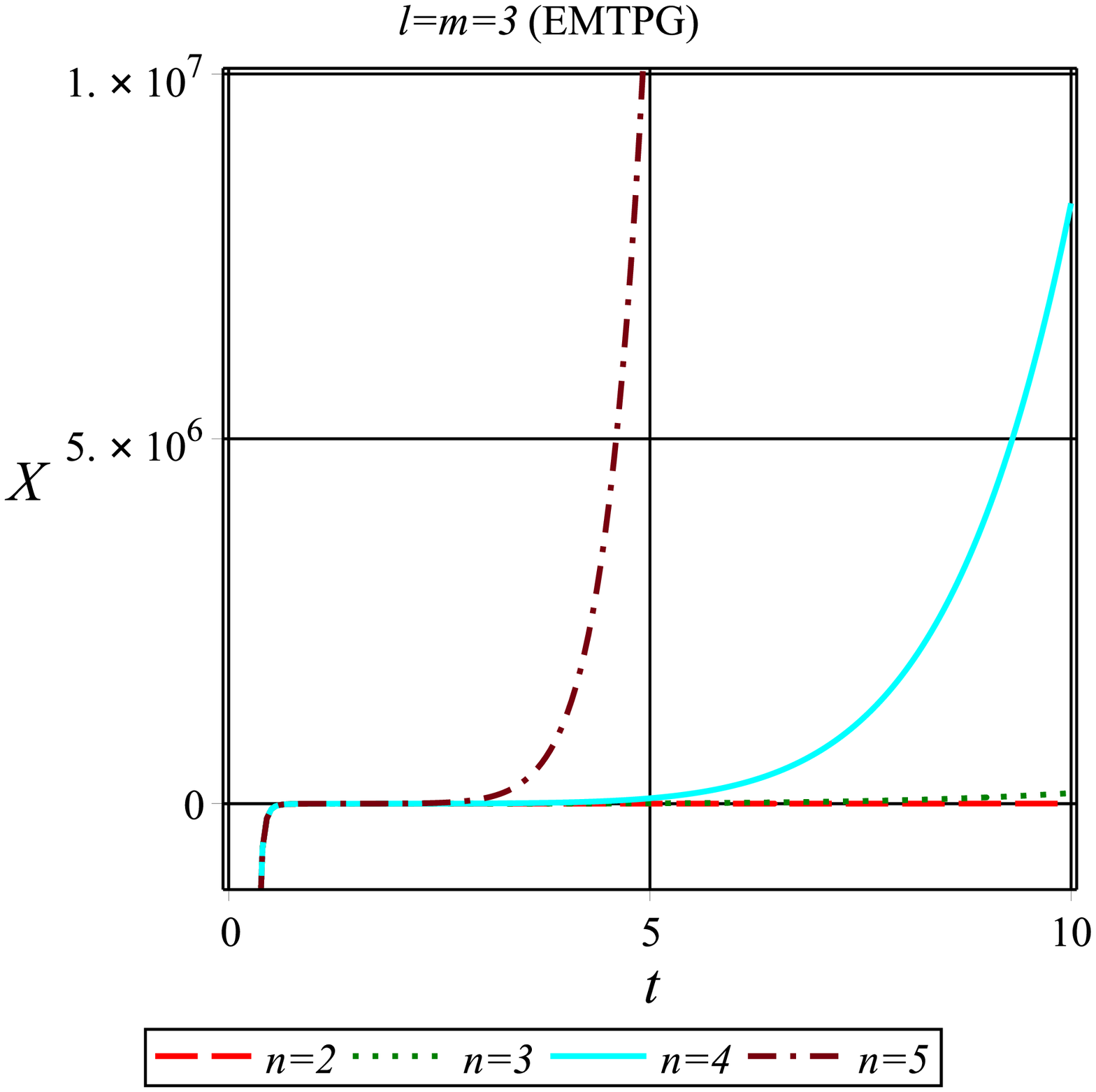}
 \end{array}$
 \end{center}
\caption{The first law of thermodynamics of the EMTPG model with $w=-\frac{1}{3}$ in unit of $G$ versus time
for $\alpha_{1}=\alpha_{2}=1$ and unit value for other constants.}
 \label{fig8}
\end{figure}

Using the relation (\ref{fT-EMTPG}) in the solution given by (\ref{cons2}), one can obtain,
\begin{equation}
z+1=\frac{1}{a}=\left[\rho^{3}\exp\left(C_{1}-\rho^{2l-1}3(\frac{4}{3})^{l}l\alpha_{2}R^{m}\right)\right]^{\frac{1}{6}}
\end{equation}
Motivated by the previous subsection, we assume the cosmic apparent horizon
radius as given by the equation (\ref{r-3}). In this case, it is
clear that $R>0$ as well as $T>0$. We know that energy density is
a decreasing function of the cosmic time and hence we assume,
\begin{equation}
\rho\propto\frac{1}{t}
\end{equation}
In that case we are able to study thermodynamics of the model
numerically. We will show that this model is stable and the first
law of thermodynamics is satisfied for a suitable choice of $n$,
$m$ and $l$. In order to check validity of the first law of
thermodynamics we use the equation (\ref{W2}). According to
Fig. \ref{fig8} we find that the first law is violated at the early
time while satisfied at the late time, if we choose suitable
values for $n$, $m$ and $l$. This is represented by the left plot
of the Fig. \ref{fig8}.

\begin{figure}[h!]
 \begin{center}$
 \begin{array}{cccc}
\includegraphics[width=70 mm]{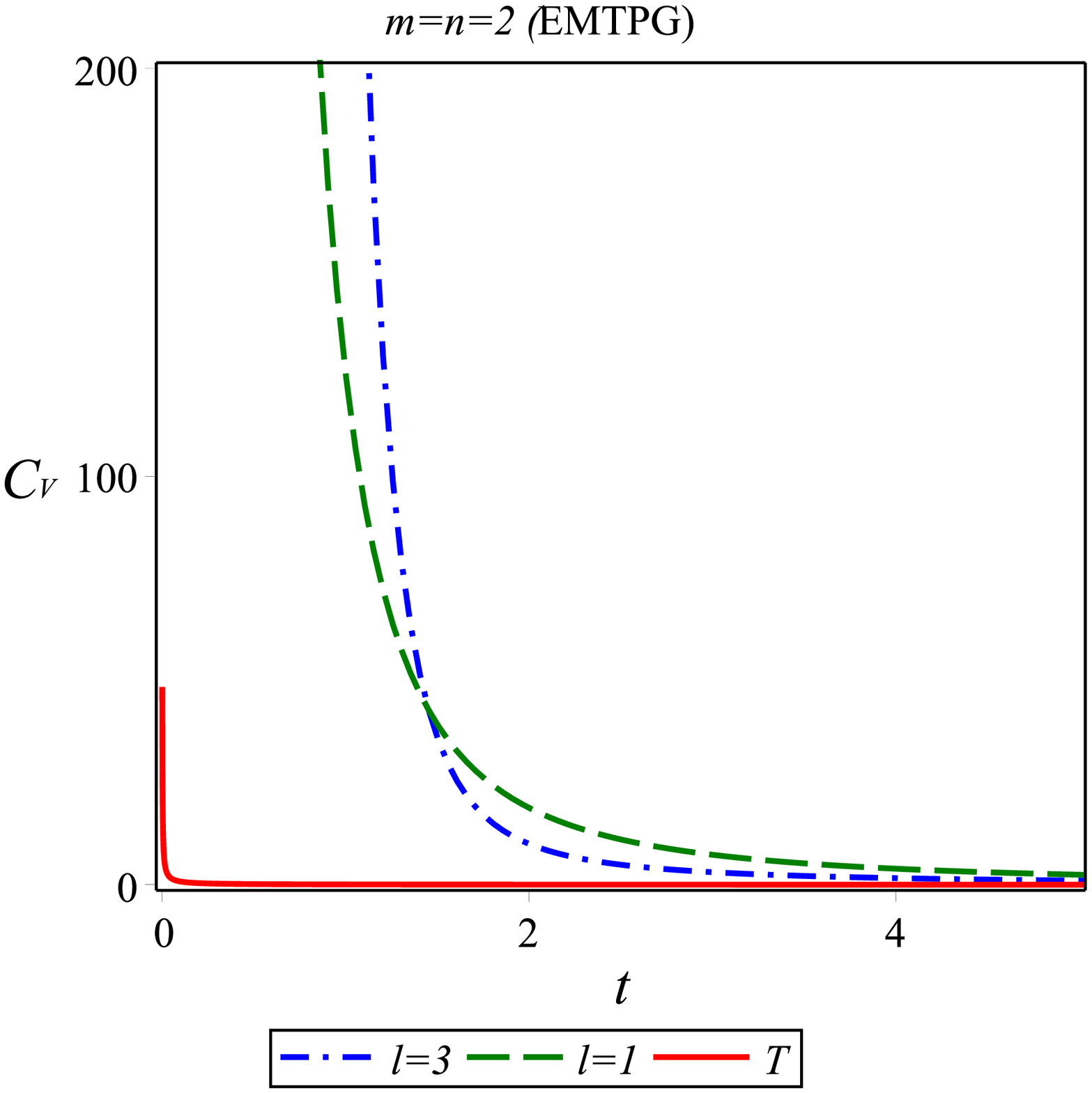}~~~~~~\includegraphics[width=70 mm]{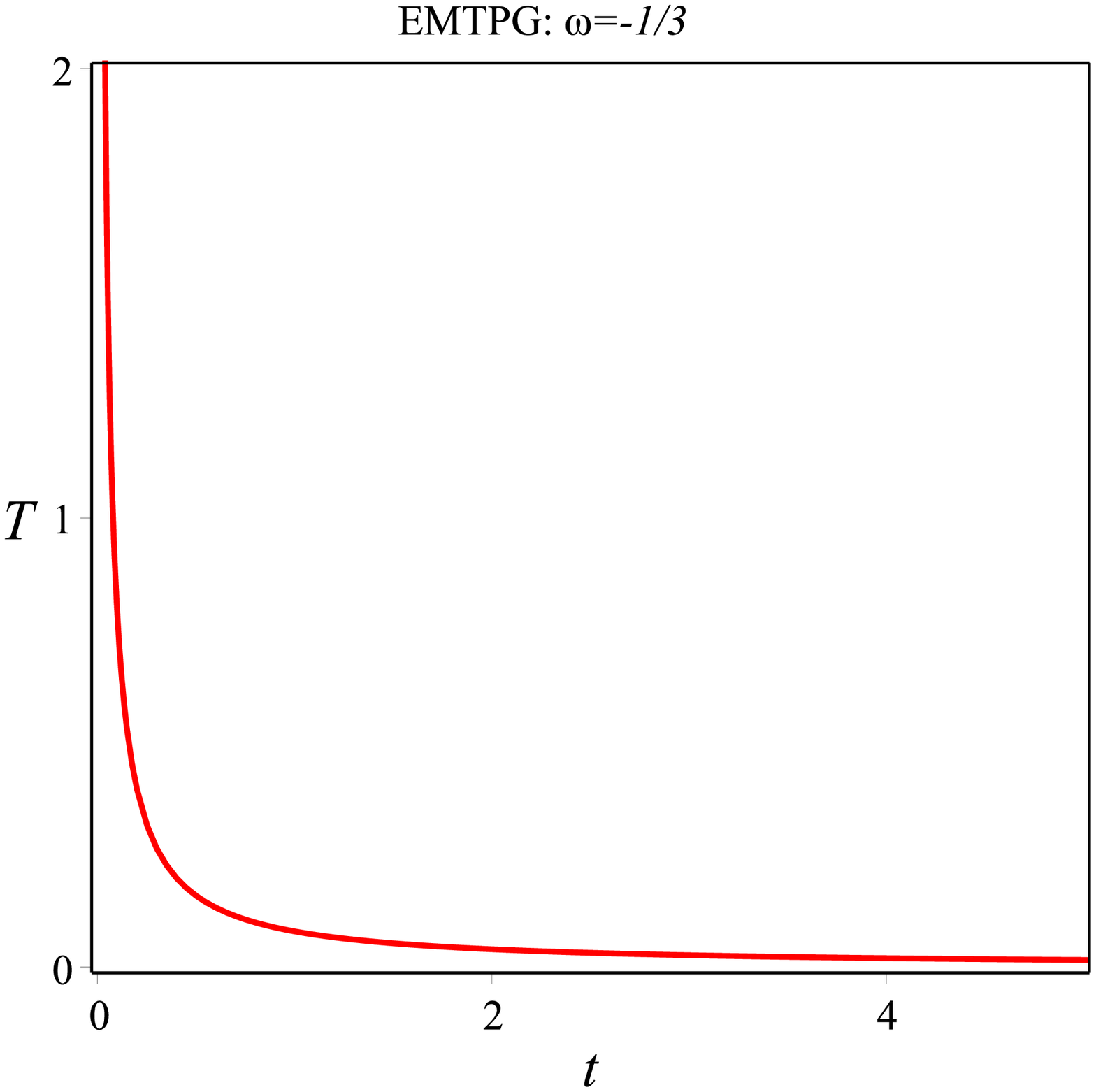}
 \end{array}$
 \end{center}
\caption{Specific heat of the EMTPG model with $w=-\frac{1}{3}$ in
unit of $G$ versus time for $\alpha_{1}=\alpha_{2}=1$ and unit
value for other constants. Temperature has also been shown in the
figure. The figure on the right shows temperature in the zoomed
range.}
 \label{fig9}
\end{figure}

Also, our numerical study on the Helmholtz free energy and
internal energy indicated that the thermodynamic potentials have a
maximum with negative value (for selected values of $l$, $m$ and
$n$ which satisfy the first law of thermodynamics) which is a sign
of the model stability. We can confirm this point by analyzing the
specific heat. We show in the Fig. \ref{fig9} that the specific
heat and temperature are positive, and hence the model may be
stable. Temperature has also been shown in a zoomed scale to get a
better idea about its nature in the figure on the right. The
specific heat has initially a higher value, which decays to an
infinitesimal constant value at the late time.

\begin{figure}[h!]
 \begin{center}$
 \begin{array}{cccc}
\includegraphics[width=80 mm]{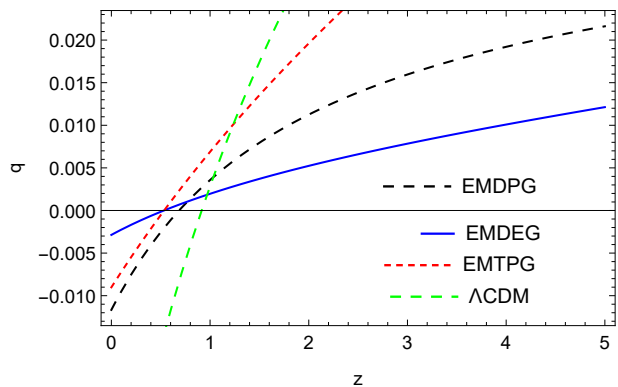}~~~~~~
 \end{array}$
 \end{center}
\caption{The deceleration parameter has been plotted against
redshift for the different models.}
 \label{fig10}
\end{figure}

In Fig. \ref{fig10} we have plotted the deceleration parameter
against redshift for the different models. Plot for the
$\Lambda$CDM model has also been shown so that a comparison can be
made considering it as a reference. It is seen that the
trajectories for the deceleration parameter $q$ enter the negative
region in the late time ($z<0.6$), showing accelerated expansion
of the universe. The shape of the trajectories are similar, but
there is a difference in phase in each model from the standard
$\Lambda$CDM model. These deviations are expected due to the
modifications of gravity.

\section{Conclusion}
In this work we have explored the thermodynamic properties of
universe in the background of the energy-momentum-squared gravity.
We reviewed the field equations of the EMSG gravity theory and
solved the non-standard continuity equation to get the expression
for the energy density $\rho$ of matter. It is found that the
continuity equation is generically integrable for only two values
of the equation of state $w=-1, -1/3$. We obtained reasonably
non-trivial expression for the energy density corresponding to
$w=-1/3$ and relatively trivial expressions for $w=-1$. However we
have conducted our thermodynamic analysis using both the values.
Then we selected our model by considering various forms of
coupling between matter and curvature. Two different types of
models were considered based on two different types of coupling
between $R$ and $T^{2}$, namely minimal and non-minimal coupling.
Various functional forms (power law and exponential) were
considered and different toy models were constructed.
Thermodynamic studies were undertaken for each of these toy models
separately and the obtained results were discussed in detail. In
the thermodynamic study, the basic thermodynamic parameters like
the entropy, specific heat, Helmholtz free energy, etc. were
determined in terms of the cosmic apparent horizon radius, $r_{A}$
and its time derivatives. The conditions for the stability of the
model have been found using the conditions of positivity of the
specific heat $C_{V}$, temperature $T$ and the existence of a
local minima in the evolution of Helmholtz free energy $F$. Since
we have two different expressions of energy density for $w=-1/3,
-1$, we have performed the thermodynamic analysis for both the
cases separately for all the models. This gave us idea about the
thermodynamic properties of the universe in EMSG gravity for
different cosmological eras. We know that $w=-1$ corresponds to
the $\Lambda CDM$ scenario, whereas $w=-1/3$ actually represents
the thin boundary between the exotic and non-exotic matter. This
scenario is cosmologically really interesting in the sense that it
corresponds to the era where the transition from ordinary matter
to dark energy takes place. It is expected that the thermodynamic
properties of the system at this juncture would be really
fascinating and may reveal some important information about the
universe. In all the three models that we have studied we have
seen that by proper fine tuning of the parameters, stability of
the model can be achieved. Obviously various parameters needed to
be constrained considerably to attain this. This is obviously
because the laws of thermodynamics had to be fulfilled and the
other stability conditions needed to be satisfied, which was not
possible over a large part of the domain. There have been various
works where the model parameters of EMSG were constrained using
observational data sets \cite{akarsu2, akarsu1, akarsu4, faria}.
In this work we were able to considerably constrain the parameter
space from the thermodynamic point of view of the system. We think
that the correct choice of the parameter space could be made by
taking into consideration both these types of analysis, which
takes us one step closer towards finding the correct model. In
this regard this work is a significant development to the EMSG
theory of gravity.

\section*{Acknowledgments}

P.R. acknowledges the Inter University Centre for Astronomy and
Astrophysics (IUCAA), Pune, India for granting visiting
associateship. We thank the referee for his/her valuable comments
that helped us to improve the quality of the paper significantly.

\section{Appendix}

\subsection{Definition of Lambert $W$ function}
Lambert $W$ function returns the value $x$ that solves the
equation
\begin{equation}\label{lambert}
y=x e^{x}
\end{equation}
Also known as omega function it is a multi-valued function, namely
the branches of the inverse relation of the function $f(u)=ue^u$,
where $u$ is any complex number and $e^u$ is the exponential
function.
\\\\

Here we provide some calculations and expressions for some
parameters which were not provided in the body of the paper for
the convenience of the reader. The expressions for effective
energy density and pressure of each model is presented below. We
report them here because these expressions are somewhat large in
size and by doing so we preserve a good presentation of the paper.

\subsection{EMDPG}

\textbf{ For $w=-1$}:\\\\

\begin{equation}
\rho_{eff}=-\frac{\varrho_{11}}{n\alpha_{1}\varrho_{1}^{n-1}},
\end{equation}
where
\begin{equation}\label{varrho 1}
\varrho_{1}=\frac{2\gamma^{2}|\varepsilon A\gamma e^{\gamma t}-2|}{(\varepsilon A\gamma e^{\gamma t}-2\varepsilon)^2},
\end{equation}
and
\begin{eqnarray}
\varrho_{11}&=&C_{01}-\frac{\alpha_{1}}{2}\varrho_{1}^{n}-\frac{4^{m}C_{01}^{2m}\alpha_{2}}{2}-\frac{3A\gamma^{3}n\alpha_{1}e^{\gamma t}\varrho_{1}^{n-1}}{\varepsilon(A\gamma e^{\gamma t}-2)^2}\nonumber\\
&-&\frac{3(n-1)n\alpha_{1}\gamma \varrho_{1}^{n-2}}{\varepsilon(A\gamma e^{\gamma t}-2)}\left[\frac{2\chi A \gamma^{4} e^{\gamma t}}{\varepsilon(A\gamma e^{\gamma t}-2)^{2}}-\frac{2A\gamma^{2}e^{\gamma t}\varrho_{1}}{A\gamma e^{\gamma t}-2}\right],
\end{eqnarray}
where $C_{01}$ given by the equation (\ref{C01}). Also, $\chi=0$ if $t=0$, $\chi=1$ if $t>\frac{\ln{\frac{2}{A\gamma\varepsilon}}}{\gamma}$, and $\chi=-1$ if $t<\frac{\ln{\frac{2}{A\gamma\varepsilon}}}{\gamma}$.\\
\begin{equation}
p_{eff}=\frac{A\gamma^{3}e^{\gamma t}}{\kappa^{2}\varepsilon(A\gamma e^{\gamma t}-2)^2}+\frac{C_{01}-\frac{1}{\kappa^{2}}(p_{11}-(n-1)n\alpha_{1}p_{12})}{n\alpha_{1}\varrho_{1}^{n-1}},
\end{equation}
where
\begin{equation}
p_{11}=\frac{\alpha_{1}}{2}\varrho_{1}^{n}+\frac{4^{m}C_{01}^{2m}\alpha_{2}}{2}+\frac{3(n-1)n\alpha_{1}\gamma \varrho_{1}^{n-2}}{\varepsilon(A\gamma e^{\gamma t}-2)}\left[\frac{2\chi A \gamma^{4} e^{\gamma t}}{\varepsilon(A\gamma e^{\gamma t}-2)^{2}}-\frac{2A\gamma^{2}e^{\gamma t}\varrho_{1}}{A\gamma e^{\gamma t}-2}\right],
\end{equation}
and
\begin{eqnarray}\label{p12}
p_{12}&=&\varrho_{1}^{n-2}\left(\frac{2\varrho_{1}A\gamma^{3}e^{\gamma t}}{A\gamma e^{\gamma t}-2}(\frac{3A\gamma e^{\gamma t}}{A\gamma e^{\gamma t}-2}-1)+\frac{2\chi A\gamma^{5}e^{\gamma t}}{\varepsilon(A\gamma e^{\gamma t}-2)}(1-\frac{4A\gamma e^{\gamma t}}{\varepsilon(A\gamma e^{\gamma t}-2)^{2}})\right)\nonumber\\
&+&\frac{n-2}{2}\varrho_{1}^{n-1}\left[\frac{2\chi A \gamma^{4} e^{\gamma t}}{\varepsilon(A\gamma e^{\gamma t}-2)^{2}}-\frac{2A\gamma^{2}e^{\gamma t}\varrho_{1}}{A\gamma e^{\gamma t}-2}\right].
\end{eqnarray}
\\\\

\textbf{ For $w=-\frac{1}{3}$}:\\\\
\begin{equation}
\rho_{eff}=-\frac{\varrho_{12}}{n\alpha_{1}\varrho_{1}^{n-1}},
\end{equation}
where $\varrho_{1}$ is given by equation (\ref{varrho 1}), while
\begin{eqnarray}
\varrho_{12}&=&\frac{\delta}{\sigma^{2}}\left(\frac{e^{\gamma t}}{A\gamma e^{\gamma t}-2}\right)^{\frac{1}{\delta}}-\frac{\alpha_{1}}{2}\varrho_{1}^{n}-\frac{\alpha_{2}}{2}\left(\frac{2\delta e^{\gamma t}}{\sigma^{2}(A\gamma e^{\gamma t}-2)}\right)^{2m}-\frac{3A\gamma^{3}n\alpha_{1}e^{\gamma t}\varrho_{1}^{n-1}}{\varepsilon(A\gamma e^{\gamma t}-2)^2}\nonumber\\
&-&\frac{3(n-1)n\alpha_{1}\gamma \varrho_{1}^{n-2}}{\varepsilon(A\gamma e^{\gamma t}-2)}\left[\frac{2\chi A \gamma^{4} e^{\gamma t}}{\varepsilon(A\gamma e^{\gamma t}-2)^{2}}-\frac{2A\gamma^{2}e^{\gamma t}\varrho_{1}}{A\gamma e^{\gamma t}-2}\right],
\end{eqnarray}
and
\begin{equation}
p_{eff}=\frac{A\gamma^{3}e^{\gamma t}}{\kappa^{2}\varepsilon(A\gamma e^{\gamma t}-2)^2}+\frac{\frac{\delta}{3\sigma^{2}e^{\frac{2\varepsilon}{\gamma}(Ae^{\gamma t}-2t)}}-\frac{1}{\kappa^{2}}(p_{13}-(n-1)n\alpha_{1}p_{12})}{n\alpha_{1}\varrho_{1}^{n-1}},
\end{equation}
where
\begin{eqnarray}
p_{13}&=&\frac{\alpha_{1}}{2}\varrho_{1}^{n}+\frac{\alpha_{2}}{2}4^{m}\left(\frac{\delta}{\sigma^{2}e^{\frac{2\varepsilon}{\gamma}(Ae^{\gamma t}-2t)}}\right)^{2m}\nonumber\\
&+&\frac{2(n-1)n\alpha_{1}\gamma \varrho_{1}^{n-2}}{\varepsilon(A\gamma e^{\gamma t}-2)}\left[\frac{2\chi A \gamma^{4} e^{\gamma t}}{\varepsilon(A\gamma e^{\gamma t}-2)^{2}}-\frac{2A\gamma^{2}e^{\gamma t}\varrho_{1}}{A\gamma e^{\gamma t}-2}\right],
\end{eqnarray}
and $p_{12}$ given by equation (\ref{p12}).

\subsection{EMDEG}
\textbf{ For $w=-1$}:\\\\
\begin{equation}
\rho_{eff}=-\frac{\varrho_{21}}{g_{1}\beta_{1}\varrho_{2}},
\end{equation}
where
\begin{equation}\label{varrho 2}
\varrho_{2}=e^{\frac{\beta_{1}\gamma^{2}|4 e^{\gamma t}-2|}{8e^{2\gamma t}}},
\end{equation}
and
\begin{eqnarray}
\varrho_{21}&=&C_{02}-\frac{g_{1}}{2}\varrho_{2}-\frac{1}{2}g_{2}e^{4\beta_{2}C_{02}^{2}}-\frac{3}{4}g_{1}\beta_{1}\gamma^{2}e^{-\gamma t}\varrho_{2}\nonumber\\
&-&\frac{3}{8}g_{1}\beta_{1}^{2}\gamma^{4}e^{-2\gamma t}\varrho_{2}\left[1-\frac{|4e^{\gamma t-2}|}{2e^{\gamma t}}\right],
\end{eqnarray}
where $C_{02}$ given by the equation (\ref{C02}).\\
\begin{equation}
p_{eff}=\frac{\gamma^{2}}{4\kappa^{2}}e^{-\gamma t}+\frac{C_{02}-\frac{1}{\kappa^{2}}\left(p_{21}+g_{1}\beta_{1}^{2}\varrho_{2}p_{22}\right)}{g_{1}\beta_{1}\varrho_{2}},
\end{equation}
where
\begin{equation}
p_{21}=\frac{1}{2}g_{1}\beta_{1}\varrho_{2}+\frac{1}{2}g_{2}e^{4\beta_{2}C_{02}^{2}}+\frac{1}{4}\gamma^{4}g_{1}\beta_{1}^{2}e^{-2\gamma t}\varrho_{2}
\left[1-\frac{|4e^{\gamma t-2}|}{2e^{\gamma t}}\right],
\end{equation}
and
\begin{eqnarray}\label{p22}
p_{22}&=&\frac{1}{4}\beta_{1}\gamma^{6}e^{-2\gamma t}\left[1-\frac{|4e^{\gamma t-2}|}{2e^{\gamma t}}\right]^{2}\nonumber\\
&+&\frac{\gamma^{4}}{2}|4e^{\gamma t-2}|e^{-2\gamma t}-\frac{3}{2}\gamma^{4}e^{-\gamma t}.
\end{eqnarray}
\\\\

\textbf{ For $w=-\frac{1}{3}$}:\\\\
Early time:\\

\begin{equation}
\rho_{eff}=-\frac{\varrho_{22}}{g_{1}\beta_{1}\varrho_{2}},
\end{equation}
where $\varrho_{2}$ given by (\ref{varrho 2}) and
\begin{eqnarray}
\varrho_{22}&=&-\frac{x_{2}g_{2}\beta_{2}}{e^{\frac{LW(x_{2})}{2}}}-\frac{g_{1}}{2}\varrho_{2}-\frac{1}{2}g_{2}e^{\frac{4\beta_{2}x_{2}^{2}}e^{LW(x_{2})}{}}
-\frac{3}{4}g_{1}\beta_{1}\gamma^{2}e^{-\gamma t}\varrho_{2}\nonumber\\
&-&\frac{3}{8}g_{1}\beta_{1}^{2}\gamma^{4}e^{-2\gamma t}\varrho_{2}\left[1-\frac{|4e^{\gamma t-2}|}{2e^{\gamma t}}\right],
\end{eqnarray}
where $LW(x_{2})$ is Lambert W function and,
\begin{equation}\label{x2}
x_{2}=-\frac{3}{8}\frac{\gamma-e^{-\gamma t}}{g_{2}^{2}\beta_{2}^{2}}.
\end{equation}

\begin{equation}
p_{eff}=\frac{\gamma^{2}}{4\kappa^{2}}e^{-\gamma t}-\frac{\frac{x_{2}g_{2}\beta_{2}}{e^{\frac{LW(x_{2})}{2}}}+\frac{1}{\kappa^{2}}\left(p_{23}+g_{1}\beta_{1}^{2}\varrho_{2}p_{22}\right)}{g_{1}\beta_{1}\varrho_{2}},
\end{equation}
where $p_{22}$ given by (\ref{p22}), while
\begin{equation}
p_{23}=\frac{1}{2}g_{1}\beta_{1}\varrho_{2}+\frac{1}{2}g_{2}e^{\frac{4x_{2}^{2}\beta_{2}}{e^{LW(x_{2})}}}+\frac{1}{4}\gamma^{4}g_{1}\beta_{1}^{2}e^{-2\gamma t}\varrho_{2}
\left[1-\frac{|4e^{\gamma t-2}|}{2e^{\gamma t}}\right].
\end{equation}

Late time:\\

\begin{equation}
\rho_{eff}=-\frac{\varrho_{23}}{g_{1}\beta_{1}\varrho_{2}},
\end{equation}
where $\varrho_{2}$ given by (\ref{varrho 2}) and
\begin{eqnarray}
\varrho_{23}&=&c_{23}e^{\frac{e^{-\gamma t}}{2}}-\frac{g_{1}}{2}\varrho_{2}-\frac{1}{2}g_{2}e^{\frac{4\beta_{2}x_{2}^{2}}e^{LW(x_{2})}{}}
-\frac{3}{4}g_{1}\beta_{1}\gamma^{2}e^{-\gamma t}\varrho_{2}\nonumber\\
&-&\frac{3}{8}g_{1}\beta_{1}^{2}\gamma^{4}e^{-2\gamma t}\varrho_{2}\left[1-\frac{|4e^{\gamma t-2}|}{2e^{\gamma t}}\right],
\end{eqnarray}
where $c_{23}$ is an integration constant.

\begin{equation}
p_{eff}=\frac{\gamma^{2}}{4\kappa^{2}}e^{-\gamma t}+\frac{c_{23}\frac{e^{-\gamma t}}{2}-\frac{1}{\kappa^{2}}\left(p_{24}+g_{1}\beta_{1}^{2}\varrho_{2}p_{22}\right)}{g_{1}\beta_{1}\varrho_{2}},
\end{equation}
where $p_{22}$ given by (\ref{p22}), while
\begin{equation}
p_{24}=\frac{1}{2}g_{1}\beta_{1}\varrho_{2}+\frac{1}{2}g_{2}\exp\left(4\beta_{2}c_{23}^{2}e^{e^{-\gamma t}}\right)+\frac{1}{4}\gamma^{4}g_{1}\beta_{1}^{2}e^{-2\gamma t}\varrho_{2}
\left[1-\frac{|4e^{\gamma t-2}|}{2e^{\gamma t}}\right].
\end{equation}
Finally, effective energy density and pressure of EMTPG model is
obtained in a similar way.


\end{document}